\documentclass[showpacs,aps,prb,reprint,superscriptaddress,preprintnumbers,nolinenumbers]{revtex4-1}

\usepackage{amsmath}
\usepackage{amssymb}
\usepackage{graphicx}
\usepackage{dcolumn}
\usepackage[dvipsnames]{xcolor}
\usepackage{amsthm,amssymb}
\usepackage{mathrsfs}
\usepackage{color}
\usepackage{endnotes}
\usepackage{bm}
\usepackage{epsfig}
\usepackage{array}
\usepackage{ulem}

\newcommand{\etal}{{\it et al.}}

\newcommand{\CsVSb}{{CsV$_{3}$Sb$_{5}$}}
 
\newcommand{\parallelsum}{\mathbin{\!/\mkern-5mu/\!}}

\begin{document}



\title{Emergence of large quantum oscillation frequencies in \\thin flakes of the kagome superconductor CsV$_{3}$Sb$_{5}$}

\author{W.~Zhang$^\S$}
\author{Lingfei~Wang$^\S$}
\author{Chun~Wai~Tsang}
\author{Xinyou~Liu}
\author{Jianyu~Xie}
\affiliation{Department of Physics, The Chinese University of Hong Kong, Shatin, Hong Kong, China}
\author{Wing~Chi~Yu}
\email[]{wingcyu@cityu.edu.hk}
\affiliation{Department of Physics, City University of Hong Kong, Kowloon, Hong Kong, China}
\author{Kwing~To~Lai}
\email[]{ktlai@phy.cuhk.edu.hk}
\affiliation{Department of Physics, The Chinese University of Hong Kong, Shatin, Hong Kong, China} 
\affiliation{Shenzhen Research Institute, The Chinese University of Hong Kong, Shatin, Hong Kong, China}
\author{Swee~K.~Goh}
\email[]{skgoh@cuhk.edu.hk}
\affiliation{Department of Physics, The Chinese University of Hong Kong, Shatin, Hong Kong, China}

\date{\today}

\begin{abstract}
Kagome metals AV$_3$Sb$_5$~(A = K, Rb, Cs) are recently discovered platforms featuring an unusual charge-density-wave (CDW) order and superconductivity. The electronic band structure of a kagome lattice can host both flat bands as well as Dirac-like bands, offering the possibility to stabilize various quantum states. Here, we probe the band structure of CsV$_3$Sb$_5$ via Shubnikov-de Haas quantum oscillations on both bulk single crystals and thin flakes. Although our frequency spectra are broadly consistent with the published data, we unambiguously reveal the existence of new frequencies with large frequencies ranging from $\sim$2085~T to $\sim$2717~T  in thin flakes when the magnetic field is along the $c$-axis. These quasi-two-dimensional  frequencies correspond to $\sim$52\% to 67\% of the CDW-distorted Brillouin zone volume. The Lifshitz-Kosevich analysis further uncovers surprisingly small cyclotron effective masses, of the order of $\sim$0.1~$m_e$, for these frequencies. Consequently, a large number of high-velocity carriers exists in the thin flake of CsV$_3$Sb$_5$. Comparing with our band structure calculations, we argue that an orbital-selective modification of the band structure is active. Our results provide indispensable information for understanding the fermiology of CsV$_3$Sb$_5$, paving a way for understanding how an orbital-selective mechanism can become an effective means to tune its electronic properties.

\end{abstract}

\maketitle

\section{Introduction} 
In a kagome lattice, the geometric frustration of the local moments provide a unique platform for the search of quantum spin liquid state \cite{Zhou2017}. Besides magnetism, the band structure of the kagome lattice naturally contains linearly dispersive Dirac cones, flat bands and van Hove singularities \cite{Kiesel2012,Kiesel2013,Wang2013}. While the Dirac cones feature non-trivial topological properties, the flat bands and the van Hove singularities promote the strong electronic correlation. The combination of geometric frustration, non-trivial topology and strong correlation in the kagome lattice provides an ideal platform to study exotic quantum phenomena, such as giant anomalous Hall effect in a ferromagnetic kagome metal Co$_3$Sn$_2$S$_2$ \cite{Wang2018}.

Recently, the discovery of kagome metals AV$_3$Sb$_5$ (A=K, Rb, Cs) has attracted a great deal of interest owing to the interplay among charge order, superconductivity and non-trivial topology \cite{Ortiz2019,Neupert2021}. They all show a second-order transition at $T_{\rm CDW} \sim$ 80--100~K, which is attributed to the appearance of a charge-density-wave (CDW) phase \cite{Du2021,Chen2021a,Li2021,Liang2021,Zhao2021,Liu2021,Hu2021,Uykur2021,Zhou2021,Jiang2021,Kang2022,Kang2022a,Wu2022,Lou2022}. Growing evidence was found to support the idea that the CDW phase could have an unconventional origin, including the observation of the chiral charge order \cite{Jiang2021,Wang2021b,Shumiya2021} and the possible time-reversal symmetry breaking without local moments \cite{Mielke2022,Yu2021c,Feng2021}. In addition, the appearance of an anomalous Hall effect, which is associated with the CDW transition, is also believed to support the scenario of an unconventional CDW state \cite{Yang2020,Yu2021b,Zheng2021}. 
At lower temperatures, a superconducting transition has been found in all members of AV$_3$Sb$_5$ at $T_c \sim$ 0.9--2.5~K \cite{Ortiz2020,Ortiz2021a,Yin2021}. Upon the application of pressure, a double-peak superconducting dome is observed in the phase diagrams of \CsVSb\ \cite{Yu2021,Chen2021a,Wang2021,Zhang2021} and RbV$_3$Sb$_5$ \cite{Wang2021a}, despite the fact that $T_{\rm CDW}$ is monotonically decreasing with pressure. This suggests an unusual competition between CDW and superconductivity.

To gain a deeper understanding of the CDW and superconductivity in AV$_3$Sb$_5$, a thorough knowledge of the Fermi surface is indispensable. Quantum oscillation (QO) investigations are one of the powerful tools to probe the Fermi surface and extract the quasiparticle effective mass, which have successfully revealed the Fermi surface topology of some candidates of topological semimetals \cite{Han2017,Hu2018,Rhodes2017,Hu2020}. 
To date, several QO reports already emerged in bulk single crystals of \CsVSb\ \cite{Yu2021,Yu2021b,Fu2021,Ortiz2021,Gan2021,Chen2022,Huang2022,Shrestha2022}. 
In all these reports, several low-frequency oscillations below $\sim$100~T, which correspond to small Fermi surface areas, have been consistently observed when the magnetic field ($B$) is along the crystalline $c$-axis. Furthermore, these frequencies are also found to have a low effective mass, and consequently they have been associated with Fermi surfaces originated from the Dirac cones. The non-trivial topological nature of some of these small frequencies were further confirmed via analyzing the phase of quantum oscillations~\cite{Fu2021}. Ortiz \etal\ \cite{Ortiz2021} and Shrestha \etal\ \cite{Shrestha2022} further reported a rich spectrum featuring several peaks ranging from 500~T to 2000~T when  $B\,  \parallelsum \,  c$  (hereafter termed the ``mid-frequency spectrum" for the ease of discussion), allowing the authors to discuss the impact of the CDW order on the Fermi surface topography, and to confirm the topologically nontrivial and quasi-2D nature of Fermi surfaces. 

The fact that the mid-frequency spectrum has not been detected consistently across the existing reports, despite a rather satisfactory agreement on the low-frequency spectrum, is most likely due to exceedingly weak signals from the mid-frequency oscillations. Thus, the QO endeavour of \CsVSb\ is far from completed, and boosting the QO signal strength is a route for further discovery. Arming with this central guiding principle, we study single crystals of \CsVSb\ with a large residual resistance ratio (RRR, defined as $R(300~{\rm K})$/$R(5~{\rm K})$). We further isolate thin flakes from these single crystals for detailed investigations. In this manuscript, we report the discovery of clear Shubnikov-de Haas (SdH) QO frequencies far beyond 2000~T (naturally termed ``high-frequency spectrum") in our thin flakes. The magnetic field angle dependence of these frequencies shows that they come from quasi-two dimensional (2D) Fermi surface sheets. Interestingly, these high-frequency peaks are found to host very small effective masses on the order of $0.1~m_e$, where $m_e$ is the free-electron mass. Thus, our results provide new fermiology insights for \CsVSb\ in a carefully designed QO experiment, and unravel the existence of a large number of high-mobility carriers that are pivotal for exotic quantum transport phenomena in \CsVSb.

\section{Methods}
\begin{figure}[!t]\centering
      \resizebox{9cm}{!}{
              \includegraphics{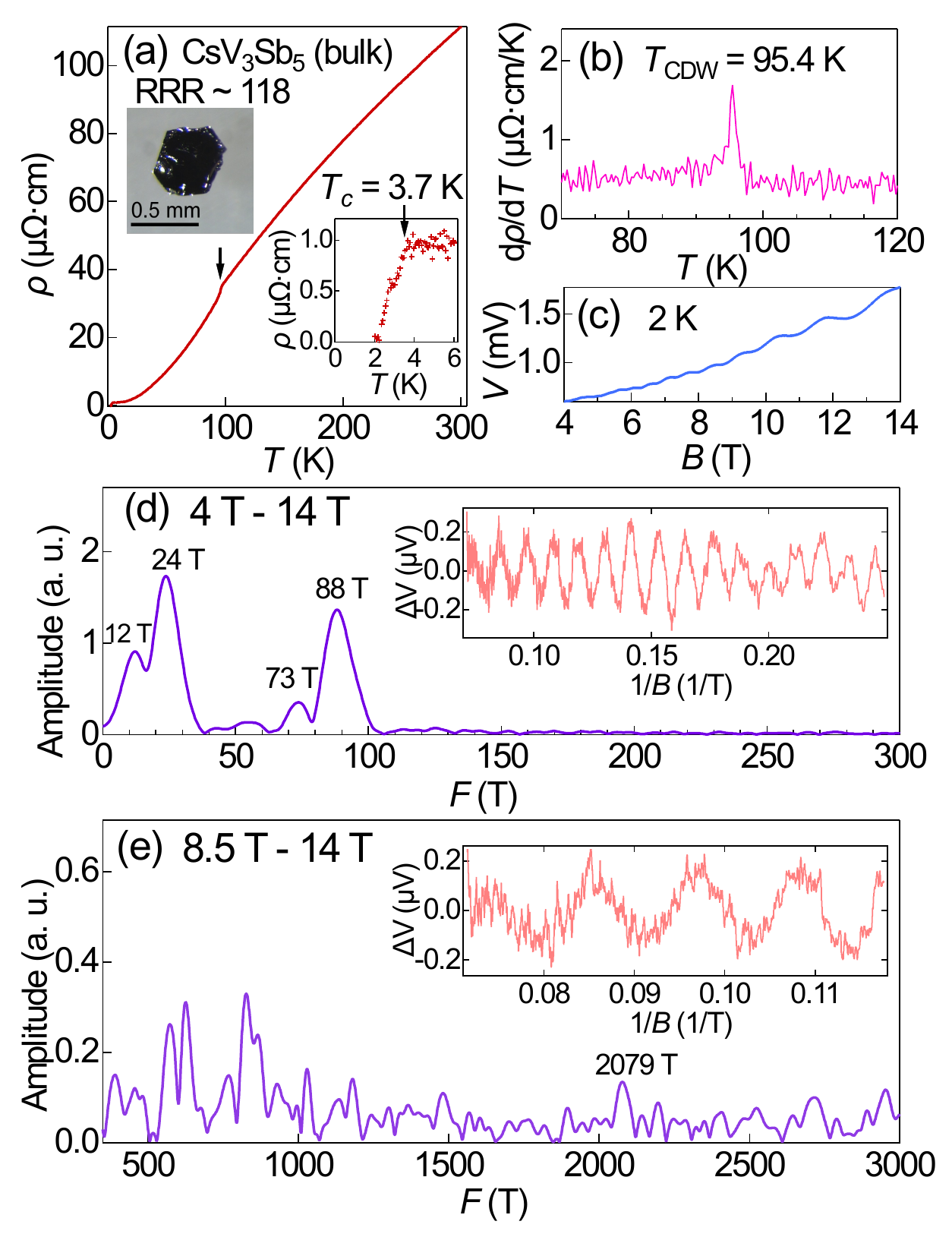}}                				
              \caption{\label{fig1}  
              (a) Temperature dependence of electrical resistivity for the bulk \CsVSb\ with a RRR of 118. The upper inset is the photograph of a typical \CsVSb\ single crystal and the lower inset shows the superconducting transition. (b) Temperature dependence of ${\rm d}\rho/{\rm d}T$, displaying a sharp peak at $T_{\rm CDW}$. (c) Raw data of the lock-in voltage versus magnetic field at 2~K.  
              (d) FFT spectrum of the bulk \CsVSb\ for the data between 4~T to 14~T. The inset shows the oscillatory signals after the removal of the background. (e) FFT spectrum of the same bulk sample for the data between 8.5~T to 14~T. The inset shows the oscillatory signals after the removal of the background.}
\end{figure}

High-quality single crystals of \CsVSb~were synthesized from Cs (ingot, 99.95 $\%$), V (powder, 99.9 $\%$) and Sb (shot, 99.9999 $\%$) using self-flux method similar to Refs.~\onlinecite{Ortiz2019,Ortiz2020}. In this study, we sealed the raw materials inside a pure-Ar-filled stainless steel jacket with the molar ratio of Cs:V:Sb = 7:3:14.
The as-grown single crystals were millimeter-sized shinny plates, as shown in the inset of Fig.~\ref{fig1}. 
Magnetotransport properties were measured by a standard four-terminal configuration in a Physical Property Measurement System by Quantum Design. 
Dupont 6838 silver paste was used for making the electrical contacts on bulk crystals, while a set of patterned electrodes was used to form a tight contact with thin flakes exfoliated from the bulk crystals~\cite{Xie2021}.
The thickness of the thin flakes was determined by a dual-beam focused ion beam system (Scios~2 DualBeam by Thermo Scientific). A Stanford Research 830 lock-in amplifier was used for QO measurements.
Density functional theory (DFT) calculations were performed, with details provided in Supplemental Material~\cite{SUPP}.

\section{Results}
\subsection{Shubnikov-de Haas quantum oscillations in \\bulk samples}
Figure~\ref{fig1}(a) shows the temperature dependence of the electrical resistivity $(\rho(T))$ for our  high-quality single-crystalline \CsVSb~in the bulk form. On cooling, $\rho(T)$ decreases and an anomaly appears at around 95~K. Correspondingly, a peak appears in  ${\rm d}\rho/{\rm d}T$, as displayed in Fig.~\ref{fig1}(b), which is consistent with the reported CDW transition. With a further cooling, $\rho(T)$ shows the onset of the superconducting transition at $\sim$3.7~K (see the lower inset in Fig.~\ref{fig1}(a)). The RRR of this sample is $\sim$118, currently one of the highest reported values. With the availability of high-quality single crystals, we further measure the magnetoresistance at the base temperature (2~K) with $B\,\parallelsum \,c$ up to 14~T. As displayed in Fig.~\ref{fig1}(c), pronounced SdH oscillations can be discerned from the raw data even without removing the background.
The fast Fourier transformation (FFT) spectrum in Fig.~\ref{fig1}(d) shows four peaks (12~T, 24~T, 73~T and 88~T) below 100~T with an excellent signal-to-noise ratio, which agree well with other groups' results~\cite{Yu2021b,Ortiz2021,Fu2021}. In addition, we also detect some weak oscillatory signals between 500~T and 2000~T, as displayed in Fig.~\ref{fig1}(e). Notably, we successfully capture one QO frequency of 2079~T, which is close to the largest frequency reported in Ref.~\onlinecite{Shrestha2022}. Similar to previous studies, the peak amplitudes in the mid-frequency spectrum are significantly less intense than those in the low-frequency spectrum.   

\begin{figure}[!t]\centering
      \resizebox{9cm}{!}{
 \includegraphics{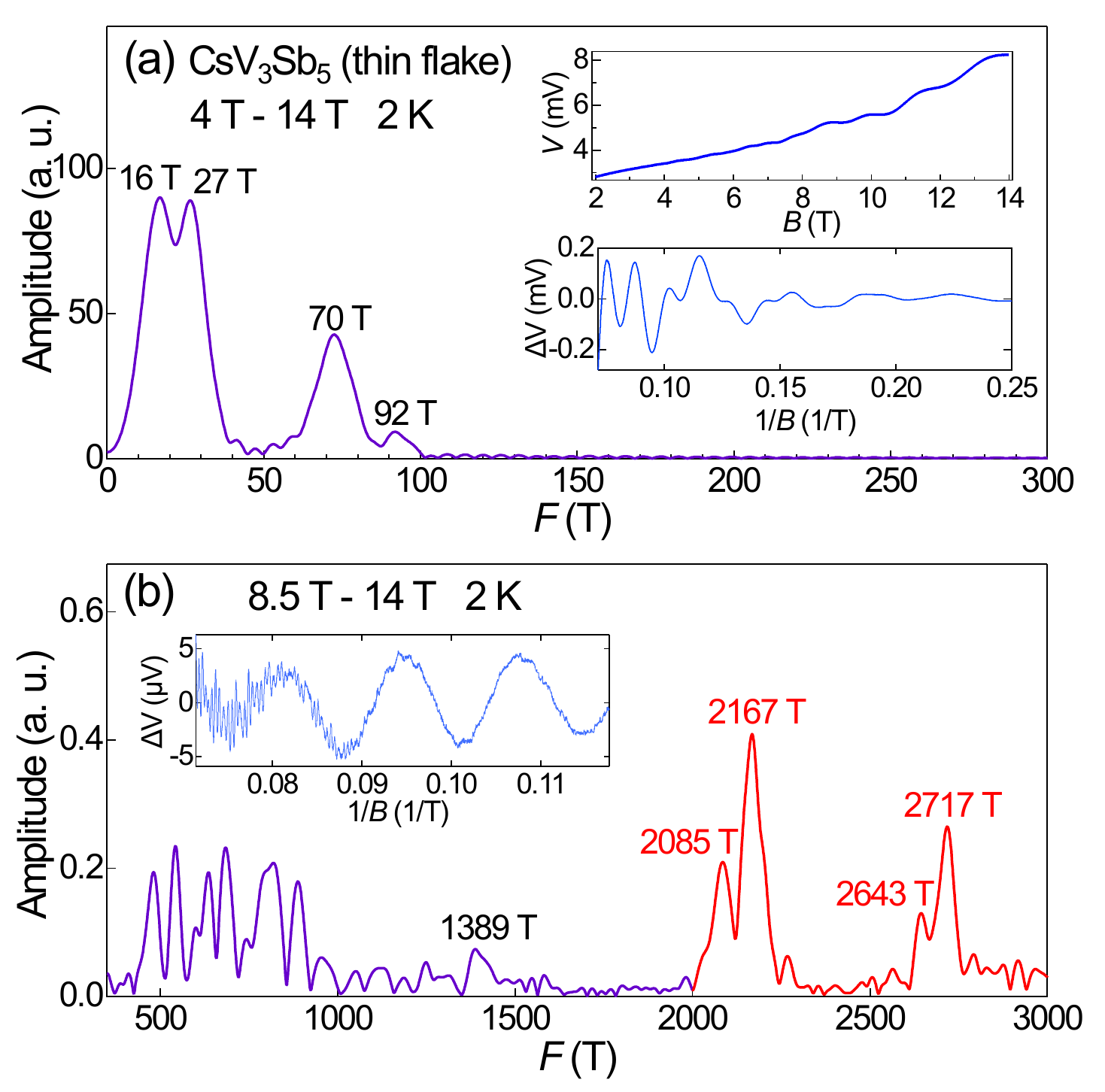}}       \caption{\label{fig2} 
(a) FFT spectrum of the thin flake (800~nm) for the data between 4~T to 14~T collected at 2~K. The upper inset is the raw data of the voltage against magnetic field. The lower inset shows the oscillatory signals after removing the background. (b) FFT spectrum of the same flake for the data between 8.5~T to 14~T. The two new groups of peaks discovered in this work are highlighted in red. The inset shows the oscillatory signals after removing the background.}            
\end{figure}

\subsection{Discovery of high-frequency peaks in thin flakes}
The highly conductive nature of our sample unavoidably makes the resistive signal extremely weak at low temperatures. To enhance the signal-to-noise ratio, we study a thin flake cleaved from the same bulk sample, thereby providing a more favourable geometric factor. Using a technique recently developed by some of us~\cite{Xie2021}, we successfully transfer a flake of \CsVSb\ with a thickness of 800~nm onto a diamond substrate pre-patterned with electrodes. The diamond substrate provides an ideal platform to ensure that the flake is thermally anchored to the cold head. As presented in Supplemental Material \cite{SUPP}, both the CDW and sharp superconducting transitions can be observed in the thin flake, but $T_{\rm{CDW}}$ decreases to around 84~K and $T_c$ increases to $\sim$4.6~K. This is probably because the flake experiences a strain from the substrate, or more intriguingly, the thin flake's carrier concentration has been modified due to air exposure \cite{Song2021}.

We conduct the magnetoresistance measurement for the thin flake of \CsVSb\ and the upper inset of Fig.~\ref{fig2}(a) shows the voltage recorded by the lock-in amplifier against the magnetic field at 2~K. The magnetic field is parallel to the $c$-axis. Figure~\ref{fig2}(a) displays the FFT spectrum for the data between 4~T and 14~T and four peaks (at 16~T, 27~T, 70~T and 92~T) are resolved, which are consistent with the observation in our bulk crystal and the reported values in bulk crystals by other groups~\cite{Yu2021b,Ortiz2021,Fu2021}. This, once again, highlights the satisfactory consistency for the low-frequency spectrum.  Besides, we detect a group of peaks between 400~T and 1000~T and one peak of 1389~T with good signal-to-noise ratio, as shown in Fig.~\ref{fig2}(b). This mid-frequency group has also been reported earlier~\cite{Ortiz2021,Shrestha2022}. These frequencies readily have some correspondence in our DFT calculations, although their definite assignment is beyond the scope of the present work. 

\begin{figure}[!t]\centering
       \resizebox{9cm}{!}{
              \includegraphics{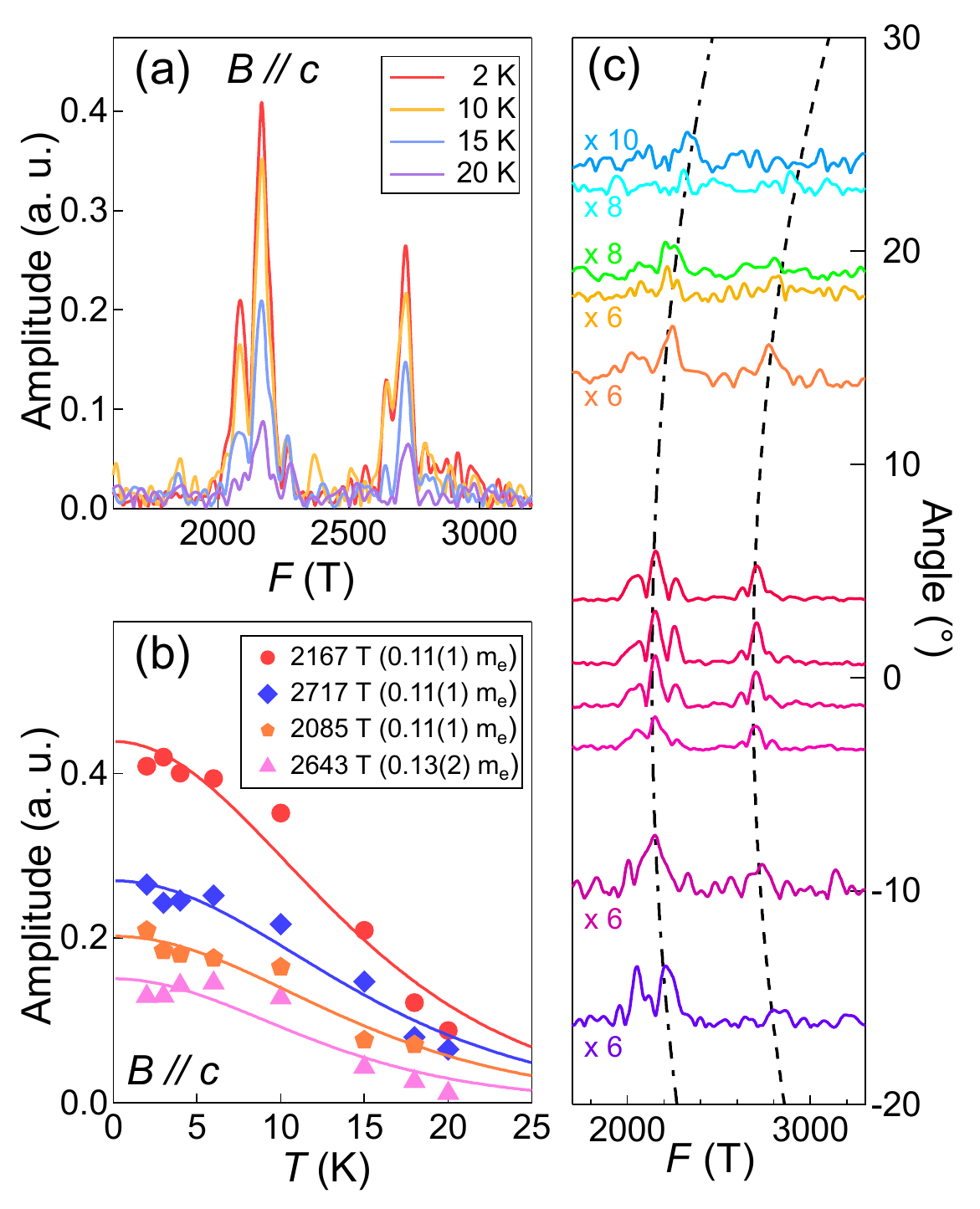}}                			
              \caption{\label{fig3}
(a) FFT spectra of SdH oscillations at different temperatures. (b) Temperature dependence of the SdH amplitudes for the peaks at 2085~T, 2167~T, 2643~T and 2717~T (solid symbols). The solid curves are fits using the thermal damping factor $R_T$. (c) FFT spectra of SdH oscillations at several field angles ($\theta$). The simulations via $F(\theta)=F(0^\circ)/\cos\theta$ are shown, with $F(0^\circ)=$ 2135~T and 2687~T for the dash-dotted and the dashed curves, respectively. $\theta=0^\circ$ corresponds to the $c$-axis, and the rotation is towards the $ab$-plane. FFT spectra at high angles have their amplitudes amplified by factors ranging from 6 to 10.}
\end{figure}

Surprisingly, we find two new groups of peaks with rather high frequencies (2085~T and 2167~T, 2643~T and 2717~T, highlighted in red in Fig.~\ref{fig2}(b)) with an excellent signal-to-noise ratio.  To the best of our knowledge, this is the first observation of QO with frequencies far beyond 2000~T in \CsVSb\ when $B\,\parallelsum \,c$, offering a precious opportunity to expand our fermiology knowledge of \CsVSb. The high-frequency spectrum shows sensitivity to both the temperature and the field angle (Fig.~\ref{fig3}). With an increasing temperature, the dominant peaks become less intense. From the thermal damping term of the Lifshitz-Kosevich theory, $R_T=X/\sinh X$ with $X=14.693m^*T/B$, the cyclotron effective mass can be extracted from the temperature dependence of the QO amplitude. As analyzed in Fig.~\ref{fig3}(b), $m^*$ for the peaks at 2085~T, 2167~T, 2643~T and 2717~T are found to be 0.11(1)~$m_e$, 0.11(1)~$m_e$, 0.13(2)~$m_e$ and 0.11(1)~$m_e$, respectively.
The relatively small $m^*$ for these four high-frequency peaks reflect the quasi-linear nature of the relevant electronic bands. Furthermore, the Fermi velocity ($v_F$) calculated from $m^*v_F=\hbar k_F$ is rather large. Approximating the orbits as circles with radii $k_F$, $v_F$ ranges from $2.53\times10^6$~m/s to $3.03\times10^6$~m/s for these high-frequency peaks. These values are roughly one order-of-magnitude higher than typical topological insulators~\cite{Qu2010,Taskin2011,Analytis2010,Kim2020}. 

\begin{figure}[h]\centering
              \includegraphics[width=.55\textwidth]{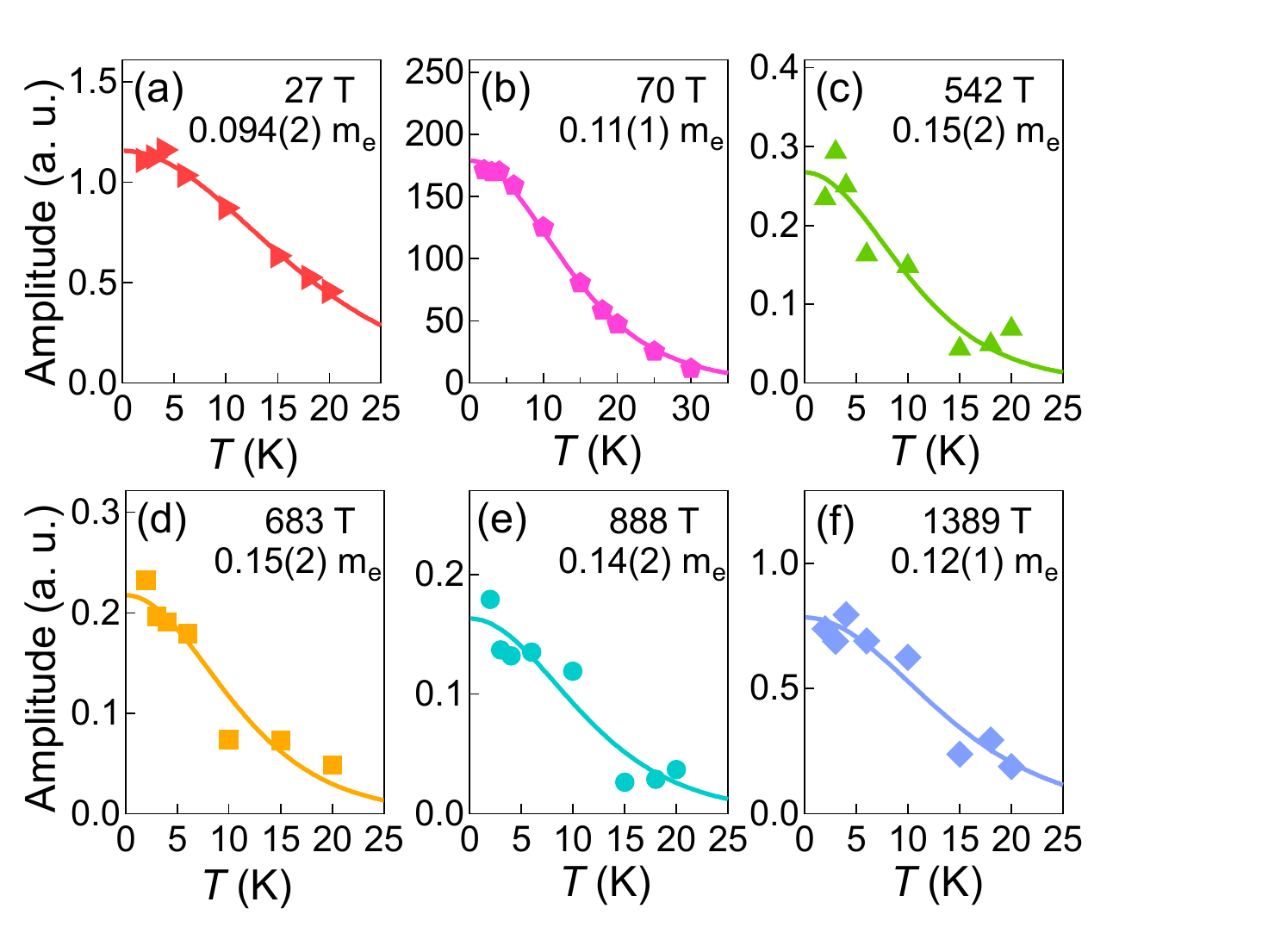}                	\caption{\label{fig4}Temperature dependence of QO amplitudes for the frequencies of 27 T, 70 T, 542 T, 683 T, 888 T and 1389 T in the thin flake of \CsVSb. The solid curves are the fit using the thermal damping factor of Lifshitz-Kosevich theory. The effective masses extracted are 0.094(2)~$m_e$ (27~T), 0.11(1)~$m_e$ (70~T), 0.15(2)~$m_e$ (542~T), 0.15(2)~$m_e$ (683~T), 0.14(2)~$m_e$ (888~T), 0.12(1)~$m_e$ (1389~T).}
              
\end{figure}

Figure~\ref{fig3}(c) shows the FFT spectra at selected magnetic field angles, focusing on the high-frequency region. When the angle between the sample $c$-axis and the magnetic field direction $(\theta)$ is large, the signal becomes weaker and therefore multiple field sweeps under identical conditions are collected to accumulate enough signals, a methodology that we successfully employed to resolve weak QO frequencies in MoTe$_2$ under pressure~\cite{Hu2020}. For instance, the FFT spectrum at $23.6^\circ$ is the result of averaging 4 sweeps. As $\theta$ increases, the QO frequencies increase according to $F(\theta)=F(0^\circ)/\cos\theta$ (see the simulations for the stronger peaks illustrated in Fig.~\ref{fig3}(c)). Such an angular dependence is a typical behaviour of quasi-2D Fermi surfaces, suggesting that the high-frequency peaks come from quasi-2D Fermi surfaces \cite{footnote1}.

\subsection{Effective masses of other peaks in thin flakes}
In addition to the high-frequency group, our relatively high signal-to-noise ratio enables the determination of the cyclotron effective mass ($m^*$) for other peaks. Figure~\ref{fig4} shows the temperature dependence of the QO amplitudes (solid symbols) for six other relatively more intense peaks. The analysis using the thermal damping factor $R_T$ (solid curves) gives the $m^*$ of $0.094(2)~m_e$, $0.11(1)~m_e$, 0.15(2)~$m_e$, 0.15(2)~$m_e$, 0.14(2)~$m_e$ and 0.12(1)~$m_e$ for 27~T, 70~T, 542~T, 683~T, 888~T and 1389~T respectively.

The $m^*$ values for the 27~T and 70~T peaks are consistent with the reported values in bulk crystals~\cite{Ortiz2021,Fu2021}. For 542~T, 683~T, 888~T and 1389~T, $m^*$ values are generally smaller than the reported values associated with comparable frequencies in Ref.~\onlinecite{Shrestha2022}.

\subsection{High-frequency peaks in other flakes} 
To prove the existence of high-frequency quantum oscillations in thin flakes, we prepare two other flakes following the same experimental procedures. The first flake ($\#$B) was cleaved from another piece of crystal from the same batch as the ones presented in Fig.~\ref{fig2}, while the second flake ($\#$C) was extracted from a crystal from an entirely different batch. The thicknesses of $\#$B and of $\#$C are 700~nm and 500~nm, respectively. 

\begin{figure}[h]\centering
              \includegraphics[width=.6\textwidth]{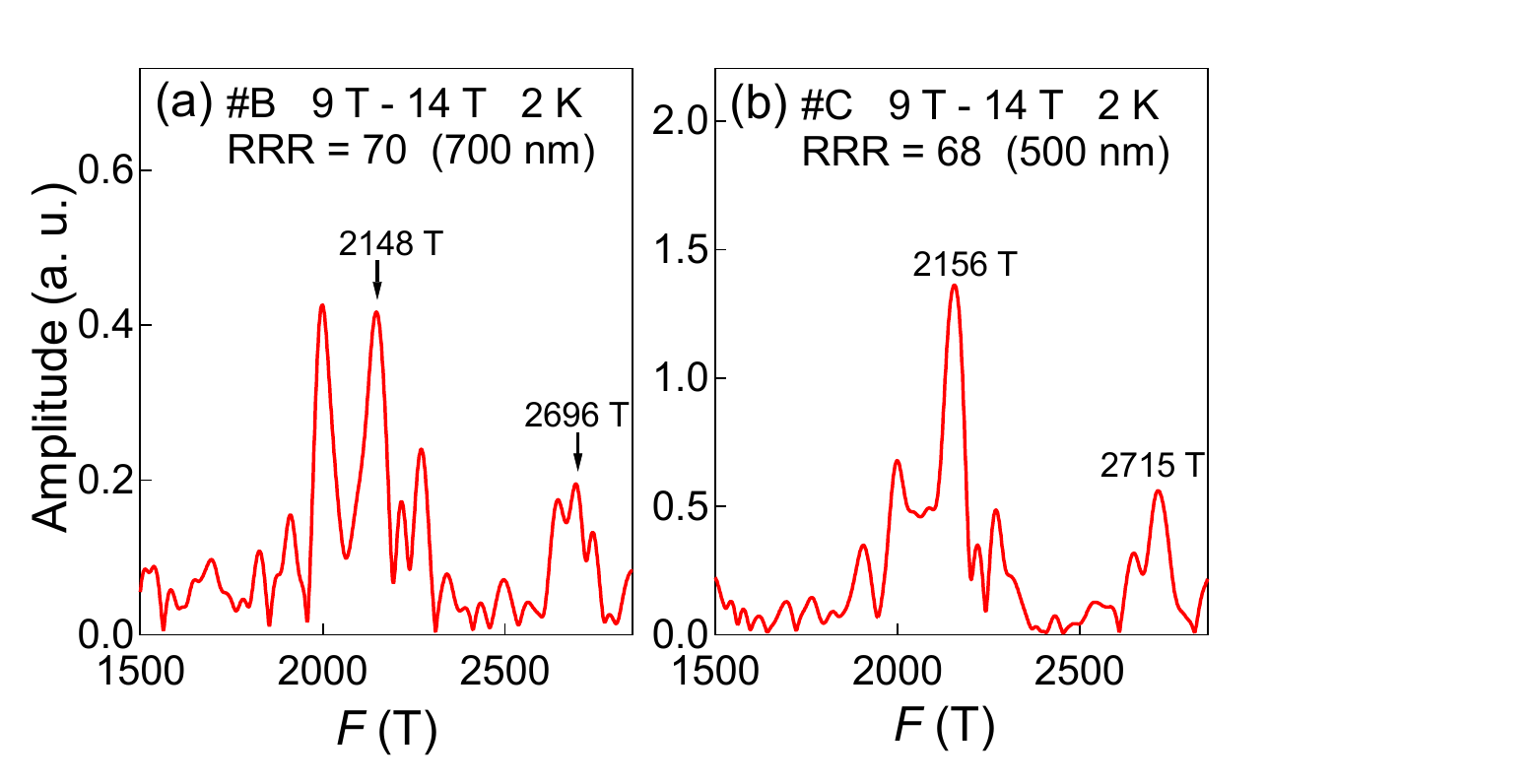}                	\caption{\label{fig5} 
               FFT spectra of SdH quantum oscillations for (a) flake $\#$B (700~nm) and (b) flake $\#$C (500~nm).
              }
\end{figure}

As shown in Figs.~\ref{fig5}(a) and \ref{fig5}(b), the FFT spectra of the new flakes are consistent with Fig.~\ref{fig2}(b) and we successfully resolve the two groups of high frequencies. The robustness of these high-frequency oscillations and our ability to reproduce them in completely different thin flakes suggest that they are the intrinsic feature.

\section{Discussion}
To understand the fermiology of \CsVSb, we perform band structure calculations.
The precise nature of the CDW state is currently under debate, but it is generally accepted that the CDW state is accompanied by an in-plane distortion which results in a $2\times2$ in-plane superstructure. Candidate distortion modes consistent with such a superstructure include a `Star of David' (SoD) and a tri-hexagonal (TrH) deformation of vanadium layers \cite{Ortiz2021,Tan2021}. X-ray diffraction, second harmonic generation measurements and Raman spectroscopy \cite{Ortiz2021,Wu2022} further revealed additional out-of-plane structural modulations, in which both TrH and SoD layers stack to increase the periodicity to 4 layers. This is commonly referred to as $2\times2\times4$ distortion.

To correctly analyze the QO spectra, band foldings must be taken into account as the increased real space periodicity introduced by the CDW order reconstructs the Fermi surfaces. We have calculated the band structure of \CsVSb\  for the $2\times2\times4$ distortion. The in-plane Brillouin zone has an area equivalent to a QO frequency of $\sim$4040~T, large enough to contain the orbits associated with the detected high-frequency peaks when $B\,\parallelsum \,c$. 
The calculated band structure is complicated, resulting in 17 Fermi surface sheets. The largest calculated frequency is associated with the quasi-2D Fermi surface sheet displayed in Fig.~\ref{fig6}(a), while the remaining sheets are presented in Supplemental Material~\cite{SUPP}. The corresponding extremal orbit is depicted in Fig.~\ref{fig6}(b). The `as-calculated' value of 1690~T for the largest frequency is still somewhat smaller than the experimental observations.

Magnetic breakdown has been shown to be active in \CsVSb~\cite{Chen2022}, and it is natural to ask if the high-frequency peaks discovered here can be attributed to magnetic breakdown. The complicated Fermi surfaces with multiple sheets in close proximity also make the system prone to magnetic breakdown. When quantum oscillation frequencies combine, the effective mass of the resultant frequency is the sum of the effective masses of the constituent components. Consider the simplest case involving only two frequencies, the effective mass of the `breakdown orbit' described by $(F_1+F_2)$ satisfies  $m^*(F_1+F_2)=m^*(F_1)+m^*(F_2)$~\cite{Shoenberg_book,Friedemann2013}. 

Given that $m^*$ of the high-frequency peaks are very similar to $m^*$ of the peaks with lower frequencies, magnetic breakdown cannot be applied here. For example, 2085~T is close to the sum of 683~T and 1389~T. However, $m^*(683~{\rm T})+m^*(1389~{\rm T})=0.27(3)~m_e$, significantly larger than $m^*(2085~{\rm T})=0.11(1)~m_e$. With the identical argument, we can also rule out the possibility that the high-frequency oscillations are harmonics.

We point out that the true value of the Fermi energy ($E_F$) can be different from the obtained value in the calculations, and it has been pointed out that the QO frequencies are very sensitive to $E_F$~\cite{Ortiz2021}. Hence, to reconcile the mismatch between the experimental and the calculated frequencies, we investigate the mechanism that can affect $E_F$ of our sample.

\begin{figure}[!t]\centering
      \resizebox{9cm}{!}{
 \includegraphics{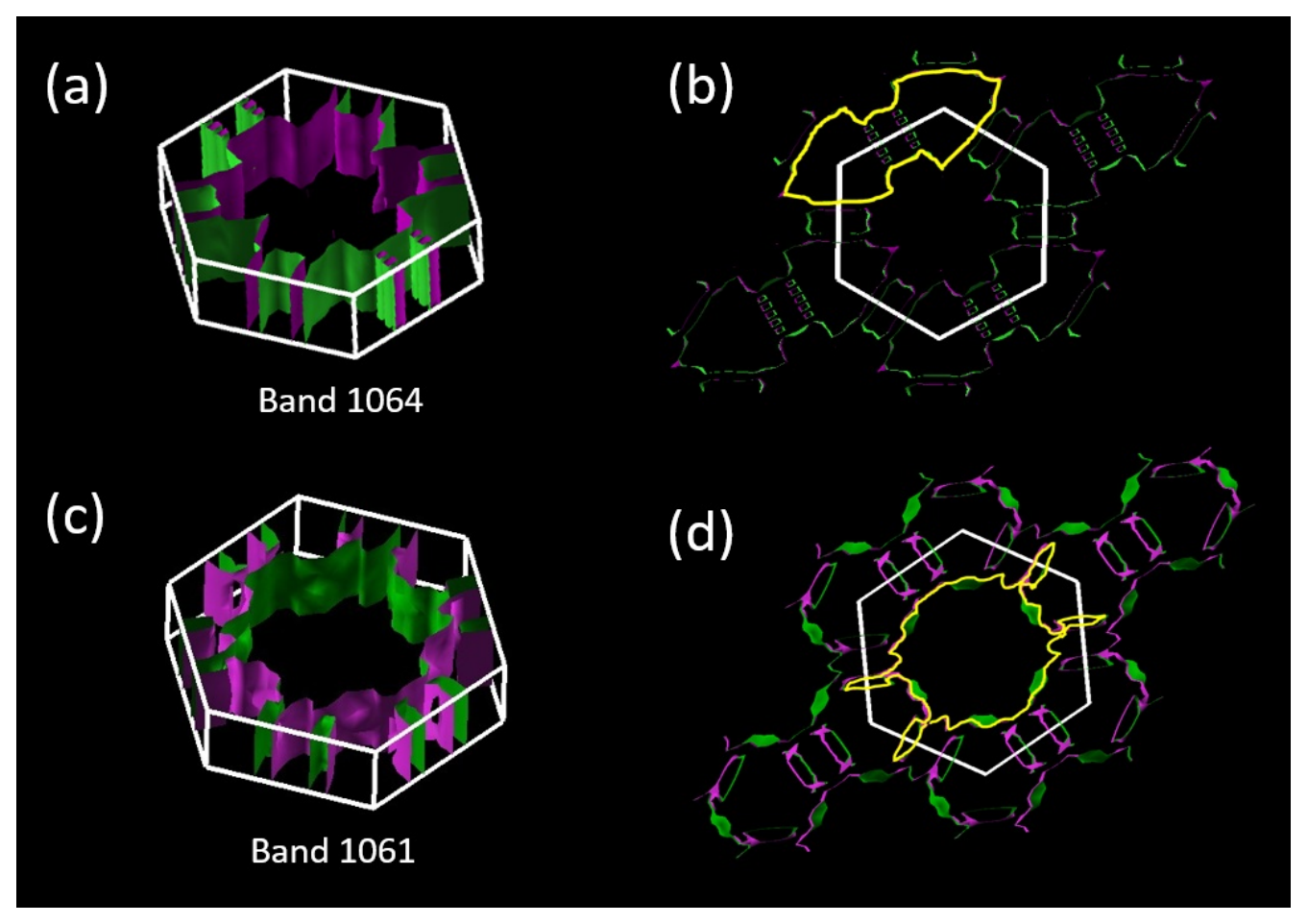}}       \caption{\label{fig6} 
(a,b) The Fermi surface sheet that hosts the largest extremal orbit when $B\,\parallelsum \, c$. The orbit is shown in yellow, which gives a QO frequency of 1690~T. The white border shows the Brillouin zone boundary. (c,d) After shifting $E_F$ downward, another Fermi surface sheet hosts the largest extremal orbit when $B\,\parallelsum \, c$. The orbit corresponds to a QO frequency of 1940~T. The illustrations in (b) and (d) are viewed along the $k_z$-axis.}            
\end{figure}

The usage of a thin flake to boost the signal-to-noise ratio offers a clue. In the scenario of `orbital-selective hole doping' proposed recently \cite{Song2021,Luo2021}, the surface of \CsVSb\ can be slightly oxidized when exposed to air. The net result is the hole doping of certain orbitals, leading to a relative downward shift of $E_F$ with respect to the bands formed by the overlap of these orbitals. Therefore, orbital-selective mechanism can lead to the increase of some QO frequencies. The effect is more pronounced in thinner samples because of the increasing surface-to-volume ratio upon thinning. It is imperative to note that not all orbitals are affected by the shift (hence the term `orbital-selective'), potentially explaining the better agreement between the flake and the bulk crystals in the low-frequency and the mid-frequency parts of the QO spectra.  Song \etal\ shows that a Cs deficiency of 5\% can give a relative band shift on the order of 100~meV for some bands~\cite{Song2021}. As an illustration, we manually shift $E_F$ downward by 58~meV, keeping the entire band structure intact. Such an operation results in a maximum frequency of 1940~T (Fig.~\ref{fig6}(d)), hosted by another Fermi surface sheet shown in Fig.~\ref{fig6}(c). To quantitatively explain our data, the precise hole concentration and a more realistic calculation that only shifts certain bands are needed. Nevertheless, our data clearly support the picture of the orbital selective hole doping in the thin flake. The hole doping readily supports the observed higher $T_c$ in our thin flake, which is also consistent with the previous report~\cite{Song2021}. Therefore, orbital-selective carrier doping is an attractive avenue to modify the carrier concentration as well as the superconducting properties of \CsVSb.
\section{Conclusion}
To summarize,  we  have  investigated  the  fermiology  of \CsVSb\ using Shubnikov-de Haas quantum oscillations. In our thin flake, we detect new frequencies associated with large extremal orbits ranging from $\sim$2085~T to $\sim$2717~T, all with surprisingly low effective masses of $\sim$0.1~$m_e$. The large number of high-velocity carriers revealed by our study will give new insights for understanding the exotic quantum transport properties in \CsVSb. Combining with the DFT calculations, we consider the $2\times2\times4$ distortion for determining the folded Fermi surface in the CDW phase. We argue that an orbital-selective modification of charge carriers can be a promising mechanism to explain the new groups of high frequency quantum oscillations. Our work not only demonstrates the rich feature of the fermiology of \CsVSb, but also supports the idea of the orbital-selective mechanism as an effective way to tune the electronic properties of kagome superconductors  AV$_3$Sb$_5$. 

{\it Note added.} When we were finalizing our manuscript, we became aware of a preprint which reports pulsed-field quantum oscillation data up to 86~T in \CsVSb~\cite{Chapai2022}. Quantum oscillation frequencies beyond 2~kT have been observed, and some of these frequencies have been attributed to magnetic breakdown~\cite{Chapai2022}. This interpretation is dissimilar to ours.


\begin{acknowledgments}

We acknowledge Wai Kin Wong for experimental support. The work was supported by Research Grants Council of Hong Kong (CUHK 14300418, CUHK 14300117, CUHK 14300419, CUHK 14301020, A-CUHK 402/19), CUHK Direct Grant (4053408, 4053461, 4053463), the National Natural Science Foundation of China (12104384), and CityU Start-up Grant (No. 9610438).

$^\S$W.Z. and L.W. contributed equally to this work.
\end{acknowledgments}


\begin{thebibliography}{62}%
\makeatletter
\providecommand \@ifxundefined [1]{%
 \@ifx{#1\undefined}
}%
\providecommand \@ifnum [1]{%
 \ifnum #1\expandafter \@firstoftwo
 \else \expandafter \@secondoftwo
 \fi
}%
\providecommand \@ifx [1]{%
 \ifx #1\expandafter \@firstoftwo
 \else \expandafter \@secondoftwo
 \fi
}%
\providecommand \natexlab [1]{#1}%
\providecommand \enquote  [1]{``#1''}%
\providecommand \bibnamefont  [1]{#1}%
\providecommand \bibfnamefont [1]{#1}%
\providecommand \citenamefont [1]{#1}%
\providecommand \href@noop [0]{\@secondoftwo}%
\providecommand \href [0]{\begingroup \@sanitize@url \@href}%
\providecommand \@href[1]{\@@startlink{#1}\@@href}%
\providecommand \@@href[1]{\endgroup#1\@@endlink}%
\providecommand \@sanitize@url [0]{\catcode `\\12\catcode `\$12\catcode
  `\&12\catcode `\#12\catcode `\^12\catcode `\_12\catcode `\%12\relax}%
\providecommand \@@startlink[1]{}%
\providecommand \@@endlink[0]{}%
\providecommand \url  [0]{\begingroup\@sanitize@url \@url }%
\providecommand \@url [1]{\endgroup\@href {#1}{\urlprefix }}%
\providecommand \urlprefix  [0]{URL }%
\providecommand \Eprint [0]{\href }%
\providecommand \doibase [0]{http://dx.doi.org/}%
\providecommand \selectlanguage [0]{\@gobble}%
\providecommand \bibinfo  [0]{\@secondoftwo}%
\providecommand \bibfield  [0]{\@secondoftwo}%
\providecommand \translation [1]{[#1]}%
\providecommand \BibitemOpen [0]{}%
\providecommand \bibitemStop [0]{}%
\providecommand \bibitemNoStop [0]{.\EOS\space}%
\providecommand \EOS [0]{\spacefactor3000\relax}%
\providecommand \BibitemShut  [1]{\csname bibitem#1\endcsname}%
\let\auto@bib@innerbib\@empty
\bibitem [{\citenamefont {Zhou}\ \emph {et~al.}(2017)\citenamefont {Zhou},
  \citenamefont {Kanoda},\ and\ \citenamefont {Ng}}]{Zhou2017}%
  \BibitemOpen
  \bibfield  {author} {\bibinfo {author} {\bibfnamefont {Y.}~\bibnamefont
  {Zhou}}, \bibinfo {author} {\bibfnamefont {K.}~\bibnamefont {Kanoda}}, \ and\
  \bibinfo {author} {\bibfnamefont {T.-K.}\ \bibnamefont {Ng}},\ }\bibfield
  {title} {\enquote {\bibinfo {title} {Quantum spin liquid states},}\ }\href
  {\doibase 10.1103/RevModPhys.89.025003} {\bibfield  {journal} {\bibinfo
  {journal} {Rev. Mod. Phys.}\ }\textbf {\bibinfo {volume} {89}},\ \bibinfo
  {pages} {025003} (\bibinfo {year} {2017})}\BibitemShut {NoStop}%
\bibitem [{\citenamefont {Kiesel}\ and\ \citenamefont
  {Thomale}(2012)}]{Kiesel2012}%
  \BibitemOpen
  \bibfield  {author} {\bibinfo {author} {\bibfnamefont {M.~L.}\ \bibnamefont
  {Kiesel}}\ and\ \bibinfo {author} {\bibfnamefont {R.}~\bibnamefont
  {Thomale}},\ }\bibfield  {title} {\enquote {\bibinfo {title} {Sublattice
  interference in the kagome {Hubbard} model},}\ }\href {\doibase
  10.1103/PhysRevB.86.121105} {\bibfield  {journal} {\bibinfo  {journal} {Phys.
  Rev. B}\ }\textbf {\bibinfo {volume} {86}},\ \bibinfo {pages} {121105}
  (\bibinfo {year} {2012})}\BibitemShut {NoStop}%
\bibitem [{\citenamefont {Kiesel}\ \emph {et~al.}(2013)\citenamefont {Kiesel},
  \citenamefont {Platt},\ and\ \citenamefont {Thomale}}]{Kiesel2013}%
  \BibitemOpen
  \bibfield  {author} {\bibinfo {author} {\bibfnamefont {M.~L.}\ \bibnamefont
  {Kiesel}}, \bibinfo {author} {\bibfnamefont {C.}~\bibnamefont {Platt}}, \
  and\ \bibinfo {author} {\bibfnamefont {R.}~\bibnamefont {Thomale}},\
  }\bibfield  {title} {\enquote {\bibinfo {title} {Unconventional {Fermi}
  surface instabilities in the kagome {Hubbard} model},}\ }\href {\doibase
  10.1103/PhysRevLett.110.126405} {\bibfield  {journal} {\bibinfo  {journal}
  {Phys. Rev. Lett.}\ }\textbf {\bibinfo {volume} {110}},\ \bibinfo {pages}
  {126405} (\bibinfo {year} {2013})}\BibitemShut {NoStop}%
\bibitem [{\citenamefont {Wang}\ \emph {et~al.}(2013)\citenamefont {Wang},
  \citenamefont {Li}, \citenamefont {Xiang},\ and\ \citenamefont
  {Wang}}]{Wang2013}%
  \BibitemOpen
  \bibfield  {author} {\bibinfo {author} {\bibfnamefont {W.-S.}\ \bibnamefont
  {Wang}}, \bibinfo {author} {\bibfnamefont {Z.-Z.}\ \bibnamefont {Li}},
  \bibinfo {author} {\bibfnamefont {Y.-Y.}\ \bibnamefont {Xiang}}, \ and\
  \bibinfo {author} {\bibfnamefont {Q.-H.}\ \bibnamefont {Wang}},\ }\bibfield
  {title} {\enquote {\bibinfo {title} {Competing electronic orders on kagome
  lattices at van {Hove} filling},}\ }\href {\doibase
  10.1103/PhysRevB.87.115135} {\bibfield  {journal} {\bibinfo  {journal} {Phys.
  Rev. B}\ }\textbf {\bibinfo {volume} {87}},\ \bibinfo {pages} {115135}
  (\bibinfo {year} {2013})}\BibitemShut {NoStop}%
\bibitem [{\citenamefont {Wang}\ \emph {et~al.}(2018)\citenamefont {Wang},
  \citenamefont {Xu}, \citenamefont {Lou}, \citenamefont {Liu}, \citenamefont
  {Li}, \citenamefont {Huang}, \citenamefont {Shen}, \citenamefont {Weng},
  \citenamefont {Wang},\ and\ \citenamefont {Lei}}]{Wang2018}%
  \BibitemOpen
  \bibfield  {author} {\bibinfo {author} {\bibfnamefont {Q.}~\bibnamefont
  {Wang}}, \bibinfo {author} {\bibfnamefont {Y.}~\bibnamefont {Xu}}, \bibinfo
  {author} {\bibfnamefont {R.}~\bibnamefont {Lou}}, \bibinfo {author}
  {\bibfnamefont {Z.}~\bibnamefont {Liu}}, \bibinfo {author} {\bibfnamefont
  {M.}~\bibnamefont {Li}}, \bibinfo {author} {\bibfnamefont {Y.}~\bibnamefont
  {Huang}}, \bibinfo {author} {\bibfnamefont {D.}~\bibnamefont {Shen}},
  \bibinfo {author} {\bibfnamefont {H.}~\bibnamefont {Weng}}, \bibinfo {author}
  {\bibfnamefont {S.}~\bibnamefont {Wang}}, \ and\ \bibinfo {author}
  {\bibfnamefont {H.}~\bibnamefont {Lei}},\ }\bibfield  {title} {\enquote
  {\bibinfo {title} {Large intrinsic anomalous {Hall} effect in half-metallic
  ferromagnet {Co$_3$Sn$_2$S$_2$} with magnetic {Weyl} fermions},}\ }\href
  {\doibase 10.1038/s41467-018-06088-2} {\bibfield  {journal} {\bibinfo
  {journal} {Nat Commun.}\ }\textbf {\bibinfo {volume} {9}},\ \bibinfo {pages}
  {3681} (\bibinfo {year} {2018})}\BibitemShut {NoStop}%
\bibitem [{\citenamefont {Ortiz}\ \emph {et~al.}(2019)\citenamefont {Ortiz},
  \citenamefont {Gomes}, \citenamefont {Morey}, \citenamefont {Winiarski},
  \citenamefont {Bordelon}, \citenamefont {Mangum}, \citenamefont {Oswald},
  \citenamefont {Rodriguez-Rivera}, \citenamefont {Neilson}, \citenamefont
  {Wilson}, \citenamefont {Ertekin}, \citenamefont {McQueen},\ and\
  \citenamefont {Toberer}}]{Ortiz2019}%
  \BibitemOpen
  \bibfield  {author} {\bibinfo {author} {\bibfnamefont {B.~R.}\ \bibnamefont
  {Ortiz}}, \bibinfo {author} {\bibfnamefont {L.~C.}\ \bibnamefont {Gomes}},
  \bibinfo {author} {\bibfnamefont {J.~R.}\ \bibnamefont {Morey}}, \bibinfo
  {author} {\bibfnamefont {M.}~\bibnamefont {Winiarski}}, \bibinfo {author}
  {\bibfnamefont {M.}~\bibnamefont {Bordelon}}, \bibinfo {author}
  {\bibfnamefont {J.~S.}\ \bibnamefont {Mangum}}, \bibinfo {author}
  {\bibfnamefont {I.~W.~H.}\ \bibnamefont {Oswald}}, \bibinfo {author}
  {\bibfnamefont {J.~A.}\ \bibnamefont {Rodriguez-Rivera}}, \bibinfo {author}
  {\bibfnamefont {J.~R.}\ \bibnamefont {Neilson}}, \bibinfo {author}
  {\bibfnamefont {S.~D.}\ \bibnamefont {Wilson}}, \bibinfo {author}
  {\bibfnamefont {E.}~\bibnamefont {Ertekin}}, \bibinfo {author} {\bibfnamefont
  {T.~M.}\ \bibnamefont {McQueen}}, \ and\ \bibinfo {author} {\bibfnamefont
  {E.~S.}\ \bibnamefont {Toberer}},\ }\bibfield  {title} {\enquote {\bibinfo
  {title} {New kagome prototype materials: discovery of {KV$_3$Sb$_5$},
  {RbV$_3$Sb$_5$}, and {CsV$_3$Sb$_5$}},}\ }\href {\doibase
  10.1103/PhysRevMaterials.3.094407} {\bibfield  {journal} {\bibinfo  {journal}
  {Phys. Rev. Mater.}\ }\textbf {\bibinfo {volume} {3}},\ \bibinfo {pages}
  {094407} (\bibinfo {year} {2019})}\BibitemShut {NoStop}%
\bibitem [{\citenamefont {Neupert}\ \emph {et~al.}(2021)\citenamefont
  {Neupert}, \citenamefont {Denner}, \citenamefont {Yin}, \citenamefont
  {Thomale},\ and\ \citenamefont {Hasan}}]{Neupert2021}%
  \BibitemOpen
  \bibfield  {author} {\bibinfo {author} {\bibfnamefont {T.}~\bibnamefont
  {Neupert}}, \bibinfo {author} {\bibfnamefont {M.~M.}\ \bibnamefont {Denner}},
  \bibinfo {author} {\bibfnamefont {J.-X.}\ \bibnamefont {Yin}}, \bibinfo
  {author} {\bibfnamefont {R.}~\bibnamefont {Thomale}}, \ and\ \bibinfo
  {author} {\bibfnamefont {M.~Z.}\ \bibnamefont {Hasan}},\ }\bibfield  {title}
  {\enquote {\bibinfo {title} {Charge order and superconductivity in kagome
  materials},}\ }\href {\doibase 10.1038/s41567-021-01404-y} {\bibfield
  {journal} {\bibinfo  {journal} {Nat. Phys.}\ }\textbf {\bibinfo {volume}
  {18}},\ \bibinfo {pages} {137} (\bibinfo {year} {2021})}\BibitemShut
  {NoStop}%
\bibitem [{\citenamefont {Du}\ \emph {et~al.}(2021)\citenamefont {Du},
  \citenamefont {Luo}, \citenamefont {Ortiz}, \citenamefont {Chen},
  \citenamefont {Duan}, \citenamefont {Zhang}, \citenamefont {Lu},
  \citenamefont {Wilson}, \citenamefont {Song},\ and\ \citenamefont
  {Yuan}}]{Du2021}%
  \BibitemOpen
  \bibfield  {author} {\bibinfo {author} {\bibfnamefont {F.}~\bibnamefont
  {Du}}, \bibinfo {author} {\bibfnamefont {S.}~\bibnamefont {Luo}}, \bibinfo
  {author} {\bibfnamefont {B.~R.}\ \bibnamefont {Ortiz}}, \bibinfo {author}
  {\bibfnamefont {Y.}~\bibnamefont {Chen}}, \bibinfo {author} {\bibfnamefont
  {W.}~\bibnamefont {Duan}}, \bibinfo {author} {\bibfnamefont {D.}~\bibnamefont
  {Zhang}}, \bibinfo {author} {\bibfnamefont {X.}~\bibnamefont {Lu}}, \bibinfo
  {author} {\bibfnamefont {S.~D.}\ \bibnamefont {Wilson}}, \bibinfo {author}
  {\bibfnamefont {Y.}~\bibnamefont {Song}}, \ and\ \bibinfo {author}
  {\bibfnamefont {H.}~\bibnamefont {Yuan}},\ }\bibfield  {title} {\enquote
  {\bibinfo {title} {Pressure-induced double superconducting domes and charge
  instability in the kagome metal {KV$_3$Sb$_5$}},}\ }\href {\doibase
  10.1103/PhysRevB.103.L220504} {\bibfield  {journal} {\bibinfo  {journal}
  {Phys. Rev. B}\ }\textbf {\bibinfo {volume} {103}},\ \bibinfo {pages}
  {L220504} (\bibinfo {year} {2021})}\BibitemShut {NoStop}%
\bibitem [{\citenamefont {Chen}\ \emph {et~al.}(2021)\citenamefont {Chen},
  \citenamefont {Wang}, \citenamefont {Yin}, \citenamefont {Gu}, \citenamefont
  {Jiang}, \citenamefont {Tu}, \citenamefont {Gong}, \citenamefont {Uwatoko},
  \citenamefont {Sun}, \citenamefont {Lei}, \citenamefont {Hu},\ and\
  \citenamefont {Cheng}}]{Chen2021a}%
  \BibitemOpen
  \bibfield  {author} {\bibinfo {author} {\bibfnamefont {K.~Y.}\ \bibnamefont
  {Chen}}, \bibinfo {author} {\bibfnamefont {N.~N.}\ \bibnamefont {Wang}},
  \bibinfo {author} {\bibfnamefont {Q.~W.}\ \bibnamefont {Yin}}, \bibinfo
  {author} {\bibfnamefont {Y.~H.}\ \bibnamefont {Gu}}, \bibinfo {author}
  {\bibfnamefont {K.}~\bibnamefont {Jiang}}, \bibinfo {author} {\bibfnamefont
  {Z.~J.}\ \bibnamefont {Tu}}, \bibinfo {author} {\bibfnamefont {C.~S.}\
  \bibnamefont {Gong}}, \bibinfo {author} {\bibfnamefont {Y.}~\bibnamefont
  {Uwatoko}}, \bibinfo {author} {\bibfnamefont {J.~P.}\ \bibnamefont {Sun}},
  \bibinfo {author} {\bibfnamefont {H.~C.}\ \bibnamefont {Lei}}, \bibinfo
  {author} {\bibfnamefont {J.~P.}\ \bibnamefont {Hu}}, \ and\ \bibinfo {author}
  {\bibfnamefont {J.-G.}\ \bibnamefont {Cheng}},\ }\bibfield  {title} {\enquote
  {\bibinfo {title} {Double superconducting dome and triple enhancement of
  {${T}_{c}$} in the kagome superconductor {CsV$_3$Sb$_5$} under high
  pressure},}\ }\href {\doibase 10.1103/PhysRevLett.126.247001} {\bibfield
  {journal} {\bibinfo  {journal} {Phys. Rev. Lett.}\ }\textbf {\bibinfo
  {volume} {126}},\ \bibinfo {pages} {247001} (\bibinfo {year}
  {2021})}\BibitemShut {NoStop}%
\bibitem [{\citenamefont {Li}\ \emph {et~al.}(2021)\citenamefont {Li},
  \citenamefont {Zhang}, \citenamefont {Yilmaz}, \citenamefont {Pai},
  \citenamefont {Marvinney}, \citenamefont {Said}, \citenamefont {Yin},
  \citenamefont {Gong}, \citenamefont {Tu}, \citenamefont {Vescovo},
  \citenamefont {Nelson}, \citenamefont {Moore}, \citenamefont {Murakami},
  \citenamefont {Lei}, \citenamefont {Lee}, \citenamefont {Lawrie},\ and\
  \citenamefont {Miao}}]{Li2021}%
  \BibitemOpen
  \bibfield  {author} {\bibinfo {author} {\bibfnamefont {H.}~\bibnamefont
  {Li}}, \bibinfo {author} {\bibfnamefont {T.~T.}\ \bibnamefont {Zhang}},
  \bibinfo {author} {\bibfnamefont {T.}~\bibnamefont {Yilmaz}}, \bibinfo
  {author} {\bibfnamefont {Y.~Y.}\ \bibnamefont {Pai}}, \bibinfo {author}
  {\bibfnamefont {C.~E.}\ \bibnamefont {Marvinney}}, \bibinfo {author}
  {\bibfnamefont {A.}~\bibnamefont {Said}}, \bibinfo {author} {\bibfnamefont
  {Q.~W.}\ \bibnamefont {Yin}}, \bibinfo {author} {\bibfnamefont {C.~S.}\
  \bibnamefont {Gong}}, \bibinfo {author} {\bibfnamefont {Z.~J.}\ \bibnamefont
  {Tu}}, \bibinfo {author} {\bibfnamefont {E.}~\bibnamefont {Vescovo}},
  \bibinfo {author} {\bibfnamefont {C.~S.}\ \bibnamefont {Nelson}}, \bibinfo
  {author} {\bibfnamefont {R.~G.}\ \bibnamefont {Moore}}, \bibinfo {author}
  {\bibfnamefont {S.}~\bibnamefont {Murakami}}, \bibinfo {author}
  {\bibfnamefont {H.~C.}\ \bibnamefont {Lei}}, \bibinfo {author} {\bibfnamefont
  {H.~N.}\ \bibnamefont {Lee}}, \bibinfo {author} {\bibfnamefont {B.~J.}\
  \bibnamefont {Lawrie}}, \ and\ \bibinfo {author} {\bibfnamefont
  {H.}~\bibnamefont {Miao}},\ }\bibfield  {title} {\enquote {\bibinfo {title}
  {Observation of unconventional charge density wave without acoustic phonon
  anomaly in kagome superconductors {$A$V$_3$Sb$_5$ ($A=\mathrm{Rb}$, Cs)}},}\
  }\href {\doibase 10.1103/PhysRevX.11.031050} {\bibfield  {journal} {\bibinfo
  {journal} {Phys. Rev. X}\ }\textbf {\bibinfo {volume} {11}},\ \bibinfo
  {pages} {031050} (\bibinfo {year} {2021})}\BibitemShut {NoStop}%
\bibitem [{\citenamefont {Liang}\ \emph {et~al.}(2021)\citenamefont {Liang},
  \citenamefont {Hou}, \citenamefont {Zhang}, \citenamefont {Ma}, \citenamefont
  {Wu}, \citenamefont {Zhang}, \citenamefont {Yu}, \citenamefont {Ying},
  \citenamefont {Jiang}, \citenamefont {Shan}, \citenamefont {Wang},\ and\
  \citenamefont {Chen}}]{Liang2021}%
  \BibitemOpen
  \bibfield  {author} {\bibinfo {author} {\bibfnamefont {Z.}~\bibnamefont
  {Liang}}, \bibinfo {author} {\bibfnamefont {X.}~\bibnamefont {Hou}}, \bibinfo
  {author} {\bibfnamefont {F.}~\bibnamefont {Zhang}}, \bibinfo {author}
  {\bibfnamefont {W.}~\bibnamefont {Ma}}, \bibinfo {author} {\bibfnamefont
  {P.}~\bibnamefont {Wu}}, \bibinfo {author} {\bibfnamefont {Z.}~\bibnamefont
  {Zhang}}, \bibinfo {author} {\bibfnamefont {F.}~\bibnamefont {Yu}}, \bibinfo
  {author} {\bibfnamefont {J.-J.}\ \bibnamefont {Ying}}, \bibinfo {author}
  {\bibfnamefont {K.}~\bibnamefont {Jiang}}, \bibinfo {author} {\bibfnamefont
  {L.}~\bibnamefont {Shan}}, \bibinfo {author} {\bibfnamefont {Z.}~\bibnamefont
  {Wang}}, \ and\ \bibinfo {author} {\bibfnamefont {X.-H.}\ \bibnamefont
  {Chen}},\ }\bibfield  {title} {\enquote {\bibinfo {title} {Three-dimensional
  charge density wave and surface-dependent vortex-core states in a kagome
  superconductor {CsV$_3$Sb$_5$}},}\ }\href {\doibase
  10.1103/PhysRevX.11.031026} {\bibfield  {journal} {\bibinfo  {journal} {Phys.
  Rev. X}\ }\textbf {\bibinfo {volume} {11}},\ \bibinfo {pages} {031026}
  (\bibinfo {year} {2021})}\BibitemShut {NoStop}%
\bibitem [{\citenamefont {Zhao}\ \emph {et~al.}(2021)\citenamefont {Zhao},
  \citenamefont {Li}, \citenamefont {Ortiz}, \citenamefont {Teicher},
  \citenamefont {Park}, \citenamefont {Ye}, \citenamefont {Wang}, \citenamefont
  {Balents}, \citenamefont {Wilson},\ and\ \citenamefont
  {Zeljkovic}}]{Zhao2021}%
  \BibitemOpen
  \bibfield  {author} {\bibinfo {author} {\bibfnamefont {H.}~\bibnamefont
  {Zhao}}, \bibinfo {author} {\bibfnamefont {H.}~\bibnamefont {Li}}, \bibinfo
  {author} {\bibfnamefont {B.~R.}\ \bibnamefont {Ortiz}}, \bibinfo {author}
  {\bibfnamefont {S.~M.~L.}\ \bibnamefont {Teicher}}, \bibinfo {author}
  {\bibfnamefont {T.}~\bibnamefont {Park}}, \bibinfo {author} {\bibfnamefont
  {M.}~\bibnamefont {Ye}}, \bibinfo {author} {\bibfnamefont {Z.}~\bibnamefont
  {Wang}}, \bibinfo {author} {\bibfnamefont {L.}~\bibnamefont {Balents}},
  \bibinfo {author} {\bibfnamefont {S.~D.}\ \bibnamefont {Wilson}}, \ and\
  \bibinfo {author} {\bibfnamefont {I.}~\bibnamefont {Zeljkovic}},\ }\bibfield
  {title} {\enquote {\bibinfo {title} {Cascade of correlated electron states in
  the kagome superconductor {CsV$_3$Sb$_5$}},}\ }\href {\doibase
  10.1038/s41586-021-03946-w} {\bibfield  {journal} {\bibinfo  {journal}
  {Nature}\ }\textbf {\bibinfo {volume} {599}},\ \bibinfo {pages} {216}
  (\bibinfo {year} {2021})}\BibitemShut {NoStop}%
\bibitem [{\citenamefont {Liu}\ \emph {et~al.}(2021)\citenamefont {Liu},
  \citenamefont {Zhao}, \citenamefont {Yin}, \citenamefont {Gong},
  \citenamefont {Tu}, \citenamefont {Li}, \citenamefont {Song}, \citenamefont
  {Liu}, \citenamefont {Shen}, \citenamefont {Huang}, \citenamefont {Liu},
  \citenamefont {Lei},\ and\ \citenamefont {Wang}}]{Liu2021}%
  \BibitemOpen
  \bibfield  {author} {\bibinfo {author} {\bibfnamefont {Z.}~\bibnamefont
  {Liu}}, \bibinfo {author} {\bibfnamefont {N.}~\bibnamefont {Zhao}}, \bibinfo
  {author} {\bibfnamefont {Q.}~\bibnamefont {Yin}}, \bibinfo {author}
  {\bibfnamefont {C.}~\bibnamefont {Gong}}, \bibinfo {author} {\bibfnamefont
  {Z.}~\bibnamefont {Tu}}, \bibinfo {author} {\bibfnamefont {M.}~\bibnamefont
  {Li}}, \bibinfo {author} {\bibfnamefont {W.}~\bibnamefont {Song}}, \bibinfo
  {author} {\bibfnamefont {Z.}~\bibnamefont {Liu}}, \bibinfo {author}
  {\bibfnamefont {D.}~\bibnamefont {Shen}}, \bibinfo {author} {\bibfnamefont
  {Y.}~\bibnamefont {Huang}}, \bibinfo {author} {\bibfnamefont
  {K.}~\bibnamefont {Liu}}, \bibinfo {author} {\bibfnamefont {H.}~\bibnamefont
  {Lei}}, \ and\ \bibinfo {author} {\bibfnamefont {S.}~\bibnamefont {Wang}},\
  }\bibfield  {title} {\enquote {\bibinfo {title} {Charge-density-wave-induced
  bands renormalization and energy gaps in a kagome superconductor
  {RbV$_3$Sb$_5$}},}\ }\href {\doibase 10.1103/PhysRevX.11.041010} {\bibfield
  {journal} {\bibinfo  {journal} {Phys. Rev. X}\ }\textbf {\bibinfo {volume}
  {11}},\ \bibinfo {pages} {041010} (\bibinfo {year} {2021})}\BibitemShut
  {NoStop}%
\bibitem [{\citenamefont {Hu}\ \emph {et~al.}(2021)\citenamefont {Hu},
  \citenamefont {Teicher}, \citenamefont {Ortiz}, \citenamefont {Luo},
  \citenamefont {Peng}, \citenamefont {Huai}, \citenamefont {Ma}, \citenamefont
  {Plumb}, \citenamefont {Wilson}, \citenamefont {He},\ and\ \citenamefont
  {Shi}}]{Hu2021}%
  \BibitemOpen
  \bibfield  {author} {\bibinfo {author} {\bibfnamefont {Y.}~\bibnamefont
  {Hu}}, \bibinfo {author} {\bibfnamefont {S.~M.~L.}\ \bibnamefont {Teicher}},
  \bibinfo {author} {\bibfnamefont {B.~R.}\ \bibnamefont {Ortiz}}, \bibinfo
  {author} {\bibfnamefont {Y.}~\bibnamefont {Luo}}, \bibinfo {author}
  {\bibfnamefont {S.}~\bibnamefont {Peng}}, \bibinfo {author} {\bibfnamefont
  {L.}~\bibnamefont {Huai}}, \bibinfo {author} {\bibfnamefont {J.~Z.}\
  \bibnamefont {Ma}}, \bibinfo {author} {\bibfnamefont {N.~C.}\ \bibnamefont
  {Plumb}}, \bibinfo {author} {\bibfnamefont {S.~D.}\ \bibnamefont {Wilson}},
  \bibinfo {author} {\bibfnamefont {J.~F.}\ \bibnamefont {He}}, \ and\ \bibinfo
  {author} {\bibfnamefont {M.}~\bibnamefont {Shi}},\ }\bibfield  {title}
  {\enquote {\bibinfo {title} {Charge-order-assisted topological surface states
  and flat bands in the kagome superconductor {CsV$_3$Sb$_5$}},}\ }\href@noop
  {} {\bibfield  {journal} {\bibinfo  {journal} {arXiv:2104.12725}\ } (\bibinfo
  {year} {2021})}\BibitemShut {NoStop}%
\bibitem [{\citenamefont {Uykur}\ \emph {et~al.}(2022)\citenamefont {Uykur},
  \citenamefont {Ortiz}, \citenamefont {Wilson}, \citenamefont {Dressel},\ and\
  \citenamefont {Tsirlin}}]{Uykur2021}%
  \BibitemOpen
  \bibfield  {author} {\bibinfo {author} {\bibfnamefont {E.}~\bibnamefont
  {Uykur}}, \bibinfo {author} {\bibfnamefont {B.~R.}\ \bibnamefont {Ortiz}},
  \bibinfo {author} {\bibfnamefont {S.~D.}\ \bibnamefont {Wilson}}, \bibinfo
  {author} {\bibfnamefont {M.}~\bibnamefont {Dressel}}, \ and\ \bibinfo
  {author} {\bibfnamefont {A.~A.}\ \bibnamefont {Tsirlin}},\ }\bibfield
  {title} {\enquote {\bibinfo {title} {Optical detection of the density-wave
  instability in the kagome metal {KV$_3$Sb$_5$}},}\ }\href {\doibase
  https://doi.org/10.1038/s41535-021-00420-8} {\bibfield  {journal} {\bibinfo
  {journal} {npj Quantum Mater.}\ }\textbf {\bibinfo {volume} {7}},\ \bibinfo
  {pages} {16} (\bibinfo {year} {2022})}\BibitemShut {NoStop}%
\bibitem [{\citenamefont {Zhou}\ \emph {et~al.}(2021)\citenamefont {Zhou},
  \citenamefont {Li}, \citenamefont {Fan}, \citenamefont {Hao}, \citenamefont
  {Dai}, \citenamefont {Wang}, \citenamefont {Yao},\ and\ \citenamefont
  {Wen}}]{Zhou2021}%
  \BibitemOpen
  \bibfield  {author} {\bibinfo {author} {\bibfnamefont {X.}~\bibnamefont
  {Zhou}}, \bibinfo {author} {\bibfnamefont {Y.}~\bibnamefont {Li}}, \bibinfo
  {author} {\bibfnamefont {X.}~\bibnamefont {Fan}}, \bibinfo {author}
  {\bibfnamefont {J.}~\bibnamefont {Hao}}, \bibinfo {author} {\bibfnamefont
  {Y.}~\bibnamefont {Dai}}, \bibinfo {author} {\bibfnamefont {Z.}~\bibnamefont
  {Wang}}, \bibinfo {author} {\bibfnamefont {Y.}~\bibnamefont {Yao}}, \ and\
  \bibinfo {author} {\bibfnamefont {H.-H.}\ \bibnamefont {Wen}},\ }\bibfield
  {title} {\enquote {\bibinfo {title} {Origin of charge density wave in the
  kagome metal {CsV$_3$Sb$_5$} as revealed by optical spectroscopy},}\ }\href
  {\doibase 10.1103/PhysRevB.104.L041101} {\bibfield  {journal} {\bibinfo
  {journal} {Phys. Rev. B}\ }\textbf {\bibinfo {volume} {104}},\ \bibinfo
  {pages} {L041101} (\bibinfo {year} {2021})}\BibitemShut {NoStop}%
\bibitem [{\citenamefont {Jiang}\ \emph {et~al.}(2021)\citenamefont {Jiang},
  \citenamefont {Yin}, \citenamefont {Denner}, \citenamefont {Shumiya},
  \citenamefont {Ortiz}, \citenamefont {Xu}, \citenamefont {Guguchia},
  \citenamefont {He}, \citenamefont {Hossain}, \citenamefont {Liu},
  \citenamefont {Ruff}, \citenamefont {Kautzsch}, \citenamefont {Zhang},
  \citenamefont {Chang}, \citenamefont {Belopolski}, \citenamefont {Zhang},
  \citenamefont {Cochran}, \citenamefont {Multer}, \citenamefont {Litskevich},
  \citenamefont {Cheng}, \citenamefont {Yang}, \citenamefont {Wang},
  \citenamefont {Thomale}, \citenamefont {Neupert}, \citenamefont {Wilson},\
  and\ \citenamefont {Hasan}}]{Jiang2021}%
  \BibitemOpen
  \bibfield  {author} {\bibinfo {author} {\bibfnamefont {Y.-X.}\ \bibnamefont
  {Jiang}}, \bibinfo {author} {\bibfnamefont {J.-X.}\ \bibnamefont {Yin}},
  \bibinfo {author} {\bibfnamefont {M.~M.}\ \bibnamefont {Denner}}, \bibinfo
  {author} {\bibfnamefont {N.}~\bibnamefont {Shumiya}}, \bibinfo {author}
  {\bibfnamefont {B.~R.}\ \bibnamefont {Ortiz}}, \bibinfo {author}
  {\bibfnamefont {G.}~\bibnamefont {Xu}}, \bibinfo {author} {\bibfnamefont
  {Z.}~\bibnamefont {Guguchia}}, \bibinfo {author} {\bibfnamefont
  {J.}~\bibnamefont {He}}, \bibinfo {author} {\bibfnamefont {M.~S.}\
  \bibnamefont {Hossain}}, \bibinfo {author} {\bibfnamefont {X.}~\bibnamefont
  {Liu}}, \bibinfo {author} {\bibfnamefont {J.}~\bibnamefont {Ruff}}, \bibinfo
  {author} {\bibfnamefont {L.}~\bibnamefont {Kautzsch}}, \bibinfo {author}
  {\bibfnamefont {S.~S.}\ \bibnamefont {Zhang}}, \bibinfo {author}
  {\bibfnamefont {G.}~\bibnamefont {Chang}}, \bibinfo {author} {\bibfnamefont
  {I.}~\bibnamefont {Belopolski}}, \bibinfo {author} {\bibfnamefont
  {Q.}~\bibnamefont {Zhang}}, \bibinfo {author} {\bibfnamefont {T.~A.}\
  \bibnamefont {Cochran}}, \bibinfo {author} {\bibfnamefont {D.}~\bibnamefont
  {Multer}}, \bibinfo {author} {\bibfnamefont {M.}~\bibnamefont {Litskevich}},
  \bibinfo {author} {\bibfnamefont {Z.-J.}\ \bibnamefont {Cheng}}, \bibinfo
  {author} {\bibfnamefont {X.~P.}\ \bibnamefont {Yang}}, \bibinfo {author}
  {\bibfnamefont {Z.}~\bibnamefont {Wang}}, \bibinfo {author} {\bibfnamefont
  {R.}~\bibnamefont {Thomale}}, \bibinfo {author} {\bibfnamefont
  {T.}~\bibnamefont {Neupert}}, \bibinfo {author} {\bibfnamefont {S.~D.}\
  \bibnamefont {Wilson}}, \ and\ \bibinfo {author} {\bibfnamefont {M.~Z.}\
  \bibnamefont {Hasan}},\ }\bibfield  {title} {\enquote {\bibinfo {title}
  {Unconventional chiral charge order in kagome superconductor
  {KV$_3$Sb$_5$}},}\ }\href {\doibase 10.1038/s41563-021-01034-y} {\bibfield
  {journal} {\bibinfo  {journal} {Nat. Mater.}\ }\textbf {\bibinfo {volume}
  {20}},\ \bibinfo {pages} {1353} (\bibinfo {year} {2021})}\BibitemShut
  {NoStop}%
\bibitem [{\citenamefont {Kang}\ \emph
  {et~al.}(2022{\natexlab{a}})\citenamefont {Kang}, \citenamefont {Fang},
  \citenamefont {Kim}, \citenamefont {Ortiz}, \citenamefont {Ryu},
  \citenamefont {Kim}, \citenamefont {Yoo}, \citenamefont {Sangiovanni},
  \citenamefont {Di~Sante}, \citenamefont {Park}, \citenamefont {Jozwiak},
  \citenamefont {Bostwick}, \citenamefont {Rotenberg}, \citenamefont {Kaxiras},
  \citenamefont {Wilson}, \citenamefont {Park},\ and\ \citenamefont
  {Comin}}]{Kang2022}%
  \BibitemOpen
  \bibfield  {author} {\bibinfo {author} {\bibfnamefont {M.}~\bibnamefont
  {Kang}}, \bibinfo {author} {\bibfnamefont {S.}~\bibnamefont {Fang}}, \bibinfo
  {author} {\bibfnamefont {J.-K.}\ \bibnamefont {Kim}}, \bibinfo {author}
  {\bibfnamefont {B.~R.}\ \bibnamefont {Ortiz}}, \bibinfo {author}
  {\bibfnamefont {S.~H.}\ \bibnamefont {Ryu}}, \bibinfo {author} {\bibfnamefont
  {J.}~\bibnamefont {Kim}}, \bibinfo {author} {\bibfnamefont {J.}~\bibnamefont
  {Yoo}}, \bibinfo {author} {\bibfnamefont {G.}~\bibnamefont {Sangiovanni}},
  \bibinfo {author} {\bibfnamefont {D.}~\bibnamefont {Di~Sante}}, \bibinfo
  {author} {\bibfnamefont {B.-G.}\ \bibnamefont {Park}}, \bibinfo {author}
  {\bibfnamefont {C.}~\bibnamefont {Jozwiak}}, \bibinfo {author} {\bibfnamefont
  {A.}~\bibnamefont {Bostwick}}, \bibinfo {author} {\bibfnamefont
  {E.}~\bibnamefont {Rotenberg}}, \bibinfo {author} {\bibfnamefont
  {E.}~\bibnamefont {Kaxiras}}, \bibinfo {author} {\bibfnamefont {S.~D.}\
  \bibnamefont {Wilson}}, \bibinfo {author} {\bibfnamefont {J.-H.}\
  \bibnamefont {Park}}, \ and\ \bibinfo {author} {\bibfnamefont
  {R.}~\bibnamefont {Comin}},\ }\bibfield  {title} {\enquote {\bibinfo {title}
  {Twofold van {Hove} singularity and origin of charge order in topological
  kagome superconductor {CsV$_3$Sb$_5$}},}\ }\href@noop {} {\bibfield
  {journal} {\bibinfo  {journal} {Nat. Phys.}\ }\textbf {\bibinfo {volume}
  {18}},\ \bibinfo {pages} {301} (\bibinfo {year}
  {2022}{\natexlab{a}})}\BibitemShut {NoStop}%
\bibitem [{\citenamefont {Kang}\ \emph
  {et~al.}(2022{\natexlab{b}})\citenamefont {Kang}, \citenamefont {Fang},
  \citenamefont {Yoo}, \citenamefont {Ortiz}, \citenamefont {Oey},
  \citenamefont {Ryu}, \citenamefont {Kim}, \citenamefont {Jozwiak},
  \citenamefont {Bostwick}, \citenamefont {Rotenberg}, \citenamefont {Kaxiras},
  \citenamefont {Checkelsky}, \citenamefont {Wilson}, \citenamefont {Park},\
  and\ \citenamefont {Comin}}]{Kang2022a}%
  \BibitemOpen
  \bibfield  {author} {\bibinfo {author} {\bibfnamefont {M.}~\bibnamefont
  {Kang}}, \bibinfo {author} {\bibfnamefont {S.}~\bibnamefont {Fang}}, \bibinfo
  {author} {\bibfnamefont {J.}~\bibnamefont {Yoo}}, \bibinfo {author}
  {\bibfnamefont {B.~R.}\ \bibnamefont {Ortiz}}, \bibinfo {author}
  {\bibfnamefont {Y.}~\bibnamefont {Oey}}, \bibinfo {author} {\bibfnamefont
  {S.~H.}\ \bibnamefont {Ryu}}, \bibinfo {author} {\bibfnamefont
  {J.}~\bibnamefont {Kim}}, \bibinfo {author} {\bibfnamefont {C.}~\bibnamefont
  {Jozwiak}}, \bibinfo {author} {\bibfnamefont {A.}~\bibnamefont {Bostwick}},
  \bibinfo {author} {\bibfnamefont {E.}~\bibnamefont {Rotenberg}}, \bibinfo
  {author} {\bibfnamefont {E.}~\bibnamefont {Kaxiras}}, \bibinfo {author}
  {\bibfnamefont {J.}~\bibnamefont {Checkelsky}}, \bibinfo {author}
  {\bibfnamefont {S.~D.}\ \bibnamefont {Wilson}}, \bibinfo {author}
  {\bibfnamefont {J.-H.}\ \bibnamefont {Park}}, \ and\ \bibinfo {author}
  {\bibfnamefont {R.}~\bibnamefont {Comin}},\ }\bibfield  {title} {\enquote
  {\bibinfo {title} {Microscopic structure of three-dimensional charge order in
  kagome superconductor {AV$_3$Sb$_5$} and its tunability},}\ }\href@noop {}
  {\bibfield  {journal} {\bibinfo  {journal} {arXiv:2202.01902}\ } (\bibinfo
  {year} {2022}{\natexlab{b}})}\BibitemShut {NoStop}%
\bibitem [{\citenamefont {Wu}\ \emph {et~al.}(2022)\citenamefont {Wu},
  \citenamefont {Ortiz}, \citenamefont {Tan}, \citenamefont {Wilson},
  \citenamefont {Yan}, \citenamefont {Birol},\ and\ \citenamefont
  {Blumberg}}]{Wu2022}%
  \BibitemOpen
  \bibfield  {author} {\bibinfo {author} {\bibfnamefont {S.}~\bibnamefont
  {Wu}}, \bibinfo {author} {\bibfnamefont {B.~R.}\ \bibnamefont {Ortiz}},
  \bibinfo {author} {\bibfnamefont {H.}~\bibnamefont {Tan}}, \bibinfo {author}
  {\bibfnamefont {S.~D.}\ \bibnamefont {Wilson}}, \bibinfo {author}
  {\bibfnamefont {B.}~\bibnamefont {Yan}}, \bibinfo {author} {\bibfnamefont
  {T.}~\bibnamefont {Birol}}, \ and\ \bibinfo {author} {\bibfnamefont
  {G.}~\bibnamefont {Blumberg}},\ }\bibfield  {title} {\enquote {\bibinfo
  {title} {Charge density wave order in kagome metal {AV$_3$Sb$_5$ (A= Cs, Rb,
  K)}},}\ }\href@noop {} {\bibfield  {journal} {\bibinfo  {journal}
  {arXiv:2201.05188}\ } (\bibinfo {year} {2022})}\BibitemShut {NoStop}%
\bibitem [{\citenamefont {Lou}\ \emph {et~al.}(2022)\citenamefont {Lou},
  \citenamefont {Fedorov}, \citenamefont {Yin}, \citenamefont {Kuibarov},
  \citenamefont {Tu}, \citenamefont {Gong}, \citenamefont {Schwier},
  \citenamefont {B\"uchner}, \citenamefont {Lei},\ and\ \citenamefont
  {Borisenko}}]{Lou2022}%
  \BibitemOpen
  \bibfield  {author} {\bibinfo {author} {\bibfnamefont {R.}~\bibnamefont
  {Lou}}, \bibinfo {author} {\bibfnamefont {A.}~\bibnamefont {Fedorov}},
  \bibinfo {author} {\bibfnamefont {Q.}~\bibnamefont {Yin}}, \bibinfo {author}
  {\bibfnamefont {A.}~\bibnamefont {Kuibarov}}, \bibinfo {author}
  {\bibfnamefont {Z.}~\bibnamefont {Tu}}, \bibinfo {author} {\bibfnamefont
  {C.}~\bibnamefont {Gong}}, \bibinfo {author} {\bibfnamefont {E.~F.}\
  \bibnamefont {Schwier}}, \bibinfo {author} {\bibfnamefont {B.}~\bibnamefont
  {B\"uchner}}, \bibinfo {author} {\bibfnamefont {H.}~\bibnamefont {Lei}}, \
  and\ \bibinfo {author} {\bibfnamefont {S.}~\bibnamefont {Borisenko}},\
  }\bibfield  {title} {\enquote {\bibinfo {title} {Charge-density-wave-induced
  peak-dip-hump structure and the multiband superconductivity in a kagome
  superconductor {CsV$_3$Sb$_5$}},}\ }\href {\doibase
  10.1103/PhysRevLett.128.036402} {\bibfield  {journal} {\bibinfo  {journal}
  {Phys. Rev. Lett.}\ }\textbf {\bibinfo {volume} {128}},\ \bibinfo {pages}
  {036402} (\bibinfo {year} {2022})}\BibitemShut {NoStop}%
\bibitem [{\citenamefont {Wang}\ \emph
  {et~al.}(2021{\natexlab{a}})\citenamefont {Wang}, \citenamefont {Jiang},
  \citenamefont {Yin}, \citenamefont {Li}, \citenamefont {Wang}, \citenamefont
  {Huang}, \citenamefont {Shao}, \citenamefont {Liu}, \citenamefont {Zhu},
  \citenamefont {Shumiya}, \citenamefont {Hossain}, \citenamefont {Liu},
  \citenamefont {Shi}, \citenamefont {Duan}, \citenamefont {Li}, \citenamefont
  {Chang}, \citenamefont {Dai}, \citenamefont {Ye}, \citenamefont {Xu},
  \citenamefont {Wang}, \citenamefont {Zheng}, \citenamefont {Jia},
  \citenamefont {Hasan},\ and\ \citenamefont {Yao}}]{Wang2021b}%
  \BibitemOpen
  \bibfield  {author} {\bibinfo {author} {\bibfnamefont {Z.}~\bibnamefont
  {Wang}}, \bibinfo {author} {\bibfnamefont {Y.-X.}\ \bibnamefont {Jiang}},
  \bibinfo {author} {\bibfnamefont {J.-X.}\ \bibnamefont {Yin}}, \bibinfo
  {author} {\bibfnamefont {Y.}~\bibnamefont {Li}}, \bibinfo {author}
  {\bibfnamefont {G.-Y.}\ \bibnamefont {Wang}}, \bibinfo {author}
  {\bibfnamefont {H.-L.}\ \bibnamefont {Huang}}, \bibinfo {author}
  {\bibfnamefont {S.}~\bibnamefont {Shao}}, \bibinfo {author} {\bibfnamefont
  {J.}~\bibnamefont {Liu}}, \bibinfo {author} {\bibfnamefont {P.}~\bibnamefont
  {Zhu}}, \bibinfo {author} {\bibfnamefont {N.}~\bibnamefont {Shumiya}},
  \bibinfo {author} {\bibfnamefont {M.~S.}\ \bibnamefont {Hossain}}, \bibinfo
  {author} {\bibfnamefont {H.}~\bibnamefont {Liu}}, \bibinfo {author}
  {\bibfnamefont {Y.}~\bibnamefont {Shi}}, \bibinfo {author} {\bibfnamefont
  {J.}~\bibnamefont {Duan}}, \bibinfo {author} {\bibfnamefont {X.}~\bibnamefont
  {Li}}, \bibinfo {author} {\bibfnamefont {G.}~\bibnamefont {Chang}}, \bibinfo
  {author} {\bibfnamefont {P.}~\bibnamefont {Dai}}, \bibinfo {author}
  {\bibfnamefont {Z.}~\bibnamefont {Ye}}, \bibinfo {author} {\bibfnamefont
  {G.}~\bibnamefont {Xu}}, \bibinfo {author} {\bibfnamefont {Y.}~\bibnamefont
  {Wang}}, \bibinfo {author} {\bibfnamefont {H.}~\bibnamefont {Zheng}},
  \bibinfo {author} {\bibfnamefont {J.}~\bibnamefont {Jia}}, \bibinfo {author}
  {\bibfnamefont {M.~Z.}\ \bibnamefont {Hasan}}, \ and\ \bibinfo {author}
  {\bibfnamefont {Y.}~\bibnamefont {Yao}},\ }\bibfield  {title} {\enquote
  {\bibinfo {title} {Electronic nature of chiral charge order in the kagome
  superconductor {CsV$_3$Sb$_5$}},}\ }\href {\doibase
  10.1103/PhysRevB.104.075148} {\bibfield  {journal} {\bibinfo  {journal}
  {Phys. Rev. B}\ }\textbf {\bibinfo {volume} {104}},\ \bibinfo {pages}
  {075148} (\bibinfo {year} {2021}{\natexlab{a}})}\BibitemShut {NoStop}%
\bibitem [{\citenamefont {Shumiya}\ \emph {et~al.}(2021)\citenamefont
  {Shumiya}, \citenamefont {Hossain}, \citenamefont {Yin}, \citenamefont
  {Jiang}, \citenamefont {Ortiz}, \citenamefont {Liu}, \citenamefont {Shi},
  \citenamefont {Yin}, \citenamefont {Lei}, \citenamefont {Zhang},
  \citenamefont {Chang}, \citenamefont {Zhang}, \citenamefont {Cochran},
  \citenamefont {Multer}, \citenamefont {Litskevich}, \citenamefont {Cheng},
  \citenamefont {Yang}, \citenamefont {Guguchia}, \citenamefont {Wilson},\ and\
  \citenamefont {Hasan}}]{Shumiya2021}%
  \BibitemOpen
  \bibfield  {author} {\bibinfo {author} {\bibfnamefont {N.}~\bibnamefont
  {Shumiya}}, \bibinfo {author} {\bibfnamefont {M.~S.}\ \bibnamefont
  {Hossain}}, \bibinfo {author} {\bibfnamefont {J.-X.}\ \bibnamefont {Yin}},
  \bibinfo {author} {\bibfnamefont {Y.-X.}\ \bibnamefont {Jiang}}, \bibinfo
  {author} {\bibfnamefont {B.~R.}\ \bibnamefont {Ortiz}}, \bibinfo {author}
  {\bibfnamefont {H.}~\bibnamefont {Liu}}, \bibinfo {author} {\bibfnamefont
  {Y.}~\bibnamefont {Shi}}, \bibinfo {author} {\bibfnamefont {Q.}~\bibnamefont
  {Yin}}, \bibinfo {author} {\bibfnamefont {H.}~\bibnamefont {Lei}}, \bibinfo
  {author} {\bibfnamefont {S.~S.}\ \bibnamefont {Zhang}}, \bibinfo {author}
  {\bibfnamefont {G.}~\bibnamefont {Chang}}, \bibinfo {author} {\bibfnamefont
  {Q.}~\bibnamefont {Zhang}}, \bibinfo {author} {\bibfnamefont {T.~A.}\
  \bibnamefont {Cochran}}, \bibinfo {author} {\bibfnamefont {D.}~\bibnamefont
  {Multer}}, \bibinfo {author} {\bibfnamefont {M.}~\bibnamefont {Litskevich}},
  \bibinfo {author} {\bibfnamefont {Z.-J.}\ \bibnamefont {Cheng}}, \bibinfo
  {author} {\bibfnamefont {X.~P.}\ \bibnamefont {Yang}}, \bibinfo {author}
  {\bibfnamefont {Z.}~\bibnamefont {Guguchia}}, \bibinfo {author}
  {\bibfnamefont {S.~D.}\ \bibnamefont {Wilson}}, \ and\ \bibinfo {author}
  {\bibfnamefont {M.~Z.}\ \bibnamefont {Hasan}},\ }\bibfield  {title} {\enquote
  {\bibinfo {title} {Intrinsic nature of chiral charge order in the kagome
  superconductor {RbV$_3$Sb$_5$}},}\ }\href {\doibase
  10.1103/PhysRevB.104.035131} {\bibfield  {journal} {\bibinfo  {journal}
  {Phys. Rev. B}\ }\textbf {\bibinfo {volume} {104}},\ \bibinfo {pages}
  {035131} (\bibinfo {year} {2021})}\BibitemShut {NoStop}%
\bibitem [{\citenamefont {{Mielke III}}\ \emph {et~al.}(2022)\citenamefont
  {{Mielke III}}, \citenamefont {Das}, \citenamefont {Yin}, \citenamefont
  {Liu}, \citenamefont {Gupta}, \citenamefont {Jiang}, \citenamefont {Medarde},
  \citenamefont {Wu}, \citenamefont {Lei}, \citenamefont {Chang}, \citenamefont
  {Dai}, \citenamefont {Si}, \citenamefont {Miao}, \citenamefont {Thomale},
  \citenamefont {Neupert}, \citenamefont {Shi}, \citenamefont {Khasanov},
  \citenamefont {Hasan}, \citenamefont {Luetkens},\ and\ \citenamefont
  {Guguchia}}]{Mielke2022}%
  \BibitemOpen
  \bibfield  {author} {\bibinfo {author} {\bibfnamefont {C.}~\bibnamefont
  {{Mielke III}}}, \bibinfo {author} {\bibfnamefont {D.}~\bibnamefont {Das}},
  \bibinfo {author} {\bibfnamefont {J.~X.}\ \bibnamefont {Yin}}, \bibinfo
  {author} {\bibfnamefont {H.}~\bibnamefont {Liu}}, \bibinfo {author}
  {\bibfnamefont {R.}~\bibnamefont {Gupta}}, \bibinfo {author} {\bibfnamefont
  {Y.~X.}\ \bibnamefont {Jiang}}, \bibinfo {author} {\bibfnamefont
  {M.}~\bibnamefont {Medarde}}, \bibinfo {author} {\bibfnamefont
  {X.}~\bibnamefont {Wu}}, \bibinfo {author} {\bibfnamefont {H.~C.}\
  \bibnamefont {Lei}}, \bibinfo {author} {\bibfnamefont {J.}~\bibnamefont
  {Chang}}, \bibinfo {author} {\bibfnamefont {P.}~\bibnamefont {Dai}}, \bibinfo
  {author} {\bibfnamefont {Q.}~\bibnamefont {Si}}, \bibinfo {author}
  {\bibfnamefont {H.}~\bibnamefont {Miao}}, \bibinfo {author} {\bibfnamefont
  {R.}~\bibnamefont {Thomale}}, \bibinfo {author} {\bibfnamefont
  {T.}~\bibnamefont {Neupert}}, \bibinfo {author} {\bibfnamefont
  {Y.}~\bibnamefont {Shi}}, \bibinfo {author} {\bibfnamefont {R.}~\bibnamefont
  {Khasanov}}, \bibinfo {author} {\bibfnamefont {M.~Z.}\ \bibnamefont {Hasan}},
  \bibinfo {author} {\bibfnamefont {H.}~\bibnamefont {Luetkens}}, \ and\
  \bibinfo {author} {\bibfnamefont {Z.}~\bibnamefont {Guguchia}},\ }\bibfield
  {title} {\enquote {\bibinfo {title} {Time-reversal symmetry-breaking charge
  order in a kagome superconductor},}\ }\href@noop {} {\bibfield  {journal}
  {\bibinfo  {journal} {Nature}\ }\textbf {\bibinfo {volume} {602}},\ \bibinfo
  {pages} {245} (\bibinfo {year} {2022})}\BibitemShut {NoStop}%
\bibitem [{\citenamefont {Yu}\ \emph {et~al.}(2021{\natexlab{a}})\citenamefont
  {Yu}, \citenamefont {Wang}, \citenamefont {Zhang}, \citenamefont {Sander},
  \citenamefont {Ni}, \citenamefont {Lu}, \citenamefont {Ma}, \citenamefont
  {Wang}, \citenamefont {Zhao}, \citenamefont {Chen}, \citenamefont {Jiang},
  \citenamefont {Zhang}, \citenamefont {Yang}, \citenamefont {Zhou},
  \citenamefont {Dong}, \citenamefont {Johnson}, \citenamefont {Graf},
  \citenamefont {Hu}, \citenamefont {Gao},\ and\ \citenamefont
  {Zhao}}]{Yu2021c}%
  \BibitemOpen
  \bibfield  {author} {\bibinfo {author} {\bibfnamefont {L.}~\bibnamefont
  {Yu}}, \bibinfo {author} {\bibfnamefont {C.}~\bibnamefont {Wang}}, \bibinfo
  {author} {\bibfnamefont {Y.}~\bibnamefont {Zhang}}, \bibinfo {author}
  {\bibfnamefont {M.}~\bibnamefont {Sander}}, \bibinfo {author} {\bibfnamefont
  {S.}~\bibnamefont {Ni}}, \bibinfo {author} {\bibfnamefont {Z.}~\bibnamefont
  {Lu}}, \bibinfo {author} {\bibfnamefont {S.}~\bibnamefont {Ma}}, \bibinfo
  {author} {\bibfnamefont {Z.}~\bibnamefont {Wang}}, \bibinfo {author}
  {\bibfnamefont {Z.}~\bibnamefont {Zhao}}, \bibinfo {author} {\bibfnamefont
  {H.}~\bibnamefont {Chen}}, \bibinfo {author} {\bibfnamefont {K.}~\bibnamefont
  {Jiang}}, \bibinfo {author} {\bibfnamefont {Y.}~\bibnamefont {Zhang}},
  \bibinfo {author} {\bibfnamefont {H.}~\bibnamefont {Yang}}, \bibinfo {author}
  {\bibfnamefont {F.}~\bibnamefont {Zhou}}, \bibinfo {author} {\bibfnamefont
  {X.}~\bibnamefont {Dong}}, \bibinfo {author} {\bibfnamefont {S.~L.}\
  \bibnamefont {Johnson}}, \bibinfo {author} {\bibfnamefont {M.~J.}\
  \bibnamefont {Graf}}, \bibinfo {author} {\bibfnamefont {J.}~\bibnamefont
  {Hu}}, \bibinfo {author} {\bibfnamefont {H.-J.}\ \bibnamefont {Gao}}, \ and\
  \bibinfo {author} {\bibfnamefont {Z.}~\bibnamefont {Zhao}},\ }\bibfield
  {title} {\enquote {\bibinfo {title} {Evidence of a hidden flux phase in the
  topological kagome metal {CsV$_3$Sb$_5$}},}\ }\href@noop {} {\bibfield
  {journal} {\bibinfo  {journal} {arXiv:2107.10714}\ } (\bibinfo {year}
  {2021}{\natexlab{a}})}\BibitemShut {NoStop}%
\bibitem [{\citenamefont {Feng}\ \emph {et~al.}(2021)\citenamefont {Feng},
  \citenamefont {Jiang}, \citenamefont {Wang},\ and\ \citenamefont
  {Hu}}]{Feng2021}%
  \BibitemOpen
  \bibfield  {author} {\bibinfo {author} {\bibfnamefont {X.}~\bibnamefont
  {Feng}}, \bibinfo {author} {\bibfnamefont {K.}~\bibnamefont {Jiang}},
  \bibinfo {author} {\bibfnamefont {Z.}~\bibnamefont {Wang}}, \ and\ \bibinfo
  {author} {\bibfnamefont {J.}~\bibnamefont {Hu}},\ }\bibfield  {title}
  {\enquote {\bibinfo {title} {Chiral flux phase in the kagome superconductor
  {AV$_3$Sb$_5$}},}\ }\href {\doibase
  https://doi.org/10.1016/j.scib.2021.04.043} {\bibfield  {journal} {\bibinfo
  {journal} {Sci. Bull.}\ }\textbf {\bibinfo {volume} {66}},\ \bibinfo {pages}
  {1384} (\bibinfo {year} {2021})}\BibitemShut {NoStop}%
\bibitem [{\citenamefont {Yang}\ \emph {et~al.}(2020)\citenamefont {Yang},
  \citenamefont {Wang}, \citenamefont {Ortiz}, \citenamefont {Liu},
  \citenamefont {Gayles}, \citenamefont {Derunova}, \citenamefont
  {Gonzalez-Hernandez}, \citenamefont {{\v{S}}mejkal}, \citenamefont {Chen},
  \citenamefont {Parkin}, \citenamefont {Wilson}, \citenamefont {Toberer},
  \citenamefont {McQueen},\ and\ \citenamefont {Ali}}]{Yang2020}%
  \BibitemOpen
  \bibfield  {author} {\bibinfo {author} {\bibfnamefont {S.-Y.}\ \bibnamefont
  {Yang}}, \bibinfo {author} {\bibfnamefont {Y.}~\bibnamefont {Wang}}, \bibinfo
  {author} {\bibfnamefont {B.~R.}\ \bibnamefont {Ortiz}}, \bibinfo {author}
  {\bibfnamefont {D.}~\bibnamefont {Liu}}, \bibinfo {author} {\bibfnamefont
  {J.}~\bibnamefont {Gayles}}, \bibinfo {author} {\bibfnamefont
  {E.}~\bibnamefont {Derunova}}, \bibinfo {author} {\bibfnamefont
  {R.}~\bibnamefont {Gonzalez-Hernandez}}, \bibinfo {author} {\bibfnamefont
  {L.}~\bibnamefont {{\v{S}}mejkal}}, \bibinfo {author} {\bibfnamefont
  {Y.}~\bibnamefont {Chen}}, \bibinfo {author} {\bibfnamefont {S.~S.~P.}\
  \bibnamefont {Parkin}}, \bibinfo {author} {\bibfnamefont {S.~D.}\
  \bibnamefont {Wilson}}, \bibinfo {author} {\bibfnamefont {E.~S.}\
  \bibnamefont {Toberer}}, \bibinfo {author} {\bibfnamefont {T.}~\bibnamefont
  {McQueen}}, \ and\ \bibinfo {author} {\bibfnamefont {M.~N.}\ \bibnamefont
  {Ali}},\ }\bibfield  {title} {\enquote {\bibinfo {title} {Giant,
  unconventional anomalous {Hall} effect in the metallic frustrated magnet
  candidate, {KV$_3$Sb$_5$}},}\ }\href@noop {} {\bibfield  {journal} {\bibinfo
  {journal} {Sci. Adv.}\ }\textbf {\bibinfo {volume} {6}},\ \bibinfo {pages}
  {eabb6003} (\bibinfo {year} {2020})}\BibitemShut {NoStop}%
\bibitem [{\citenamefont {Yu}\ \emph {et~al.}(2021{\natexlab{b}})\citenamefont
  {Yu}, \citenamefont {Wu}, \citenamefont {Wang}, \citenamefont {Lei},
  \citenamefont {Zhuo}, \citenamefont {Ying},\ and\ \citenamefont
  {Chen}}]{Yu2021b}%
  \BibitemOpen
  \bibfield  {author} {\bibinfo {author} {\bibfnamefont {F.~H.}\ \bibnamefont
  {Yu}}, \bibinfo {author} {\bibfnamefont {T.}~\bibnamefont {Wu}}, \bibinfo
  {author} {\bibfnamefont {Z.~Y.}\ \bibnamefont {Wang}}, \bibinfo {author}
  {\bibfnamefont {B.}~\bibnamefont {Lei}}, \bibinfo {author} {\bibfnamefont
  {W.~Z.}\ \bibnamefont {Zhuo}}, \bibinfo {author} {\bibfnamefont {J.~J.}\
  \bibnamefont {Ying}}, \ and\ \bibinfo {author} {\bibfnamefont {X.~H.}\
  \bibnamefont {Chen}},\ }\bibfield  {title} {\enquote {\bibinfo {title}
  {Concurrence of anomalous {Hall} effect and charge density wave in a
  superconducting topological kagome metal},}\ }\href@noop {} {\bibfield
  {journal} {\bibinfo  {journal} {Phys. Rev. B}\ }\textbf {\bibinfo {volume}
  {104}},\ \bibinfo {pages} {L041103} (\bibinfo {year}
  {2021}{\natexlab{b}})}\BibitemShut {NoStop}%
\bibitem [{\citenamefont {Zheng}\ \emph {et~al.}(2021)\citenamefont {Zheng},
  \citenamefont {Chen}, \citenamefont {Tan}, \citenamefont {Wang},
  \citenamefont {Zhu}, \citenamefont {Albarakati}, \citenamefont {Algarni},
  \citenamefont {Partridge}, \citenamefont {Farrar}, \citenamefont {Zhou},
  \citenamefont {Ning}, \citenamefont {Tian}, \citenamefont {Fuhrer},\ and\
  \citenamefont {Wang}}]{Zheng2021}%
  \BibitemOpen
  \bibfield  {author} {\bibinfo {author} {\bibfnamefont {G.}~\bibnamefont
  {Zheng}}, \bibinfo {author} {\bibfnamefont {Z.}~\bibnamefont {Chen}},
  \bibinfo {author} {\bibfnamefont {C.}~\bibnamefont {Tan}}, \bibinfo {author}
  {\bibfnamefont {M.}~\bibnamefont {Wang}}, \bibinfo {author} {\bibfnamefont
  {X.}~\bibnamefont {Zhu}}, \bibinfo {author} {\bibfnamefont {S.}~\bibnamefont
  {Albarakati}}, \bibinfo {author} {\bibfnamefont {M.}~\bibnamefont {Algarni}},
  \bibinfo {author} {\bibfnamefont {J.}~\bibnamefont {Partridge}}, \bibinfo
  {author} {\bibfnamefont {L.}~\bibnamefont {Farrar}}, \bibinfo {author}
  {\bibfnamefont {J.}~\bibnamefont {Zhou}}, \bibinfo {author} {\bibfnamefont
  {W.}~\bibnamefont {Ning}}, \bibinfo {author} {\bibfnamefont {M.}~\bibnamefont
  {Tian}}, \bibinfo {author} {\bibfnamefont {M.~S.}\ \bibnamefont {Fuhrer}}, \
  and\ \bibinfo {author} {\bibfnamefont {L.}~\bibnamefont {Wang}},\ }\bibfield
  {title} {\enquote {\bibinfo {title} {Gate-controllable giant anomalous {Hall}
  effect from flat bands in kagome metal {CsV$_3$Sb$_5$} nanoflakes},}\
  }\href@noop {} {\bibfield  {journal} {\bibinfo  {journal} {arXiv:2109.12588}\
  } (\bibinfo {year} {2021})}\BibitemShut {NoStop}%
\bibitem [{\citenamefont {Ortiz}\ \emph {et~al.}(2020)\citenamefont {Ortiz},
  \citenamefont {Teicher}, \citenamefont {Hu}, \citenamefont {Zuo},
  \citenamefont {Sarte}, \citenamefont {Schueller}, \citenamefont {Abeykoon},
  \citenamefont {Krogstad}, \citenamefont {Rosenkranz}, \citenamefont {Osborn},
  \citenamefont {Seshadri}, \citenamefont {Balents}, \citenamefont {He},\ and\
  \citenamefont {Wilson}}]{Ortiz2020}%
  \BibitemOpen
  \bibfield  {author} {\bibinfo {author} {\bibfnamefont {B.~R.}\ \bibnamefont
  {Ortiz}}, \bibinfo {author} {\bibfnamefont {S.~M.~L.}\ \bibnamefont
  {Teicher}}, \bibinfo {author} {\bibfnamefont {Y.}~\bibnamefont {Hu}},
  \bibinfo {author} {\bibfnamefont {J.~L.}\ \bibnamefont {Zuo}}, \bibinfo
  {author} {\bibfnamefont {P.~M.}\ \bibnamefont {Sarte}}, \bibinfo {author}
  {\bibfnamefont {E.~C.}\ \bibnamefont {Schueller}}, \bibinfo {author}
  {\bibfnamefont {A.~M.~M.}\ \bibnamefont {Abeykoon}}, \bibinfo {author}
  {\bibfnamefont {M.~J.}\ \bibnamefont {Krogstad}}, \bibinfo {author}
  {\bibfnamefont {S.}~\bibnamefont {Rosenkranz}}, \bibinfo {author}
  {\bibfnamefont {R.}~\bibnamefont {Osborn}}, \bibinfo {author} {\bibfnamefont
  {R.}~\bibnamefont {Seshadri}}, \bibinfo {author} {\bibfnamefont
  {L.}~\bibnamefont {Balents}}, \bibinfo {author} {\bibfnamefont
  {J.}~\bibnamefont {He}}, \ and\ \bibinfo {author} {\bibfnamefont {S.~D.}\
  \bibnamefont {Wilson}},\ }\bibfield  {title} {\enquote {\bibinfo {title}
  {{CsV$_3$Sb$_5$}: A {${\mathbb{Z}}_{2}$} topological kagome metal with a
  superconducting ground state},}\ }\href {\doibase
  10.1103/PhysRevLett.125.247002} {\bibfield  {journal} {\bibinfo  {journal}
  {Phys. Rev. Lett.}\ }\textbf {\bibinfo {volume} {125}},\ \bibinfo {pages}
  {247002} (\bibinfo {year} {2020})}\BibitemShut {NoStop}%
\bibitem [{\citenamefont {Ortiz}\ \emph
  {et~al.}(2021{\natexlab{a}})\citenamefont {Ortiz}, \citenamefont {Sarte},
  \citenamefont {Kenney}, \citenamefont {Graf}, \citenamefont {Teicher},
  \citenamefont {Seshadri},\ and\ \citenamefont {Wilson}}]{Ortiz2021a}%
  \BibitemOpen
  \bibfield  {author} {\bibinfo {author} {\bibfnamefont {B.~R.}\ \bibnamefont
  {Ortiz}}, \bibinfo {author} {\bibfnamefont {P.~M.}\ \bibnamefont {Sarte}},
  \bibinfo {author} {\bibfnamefont {E.~M.}\ \bibnamefont {Kenney}}, \bibinfo
  {author} {\bibfnamefont {M.~J.}\ \bibnamefont {Graf}}, \bibinfo {author}
  {\bibfnamefont {S.~M.~L.}\ \bibnamefont {Teicher}}, \bibinfo {author}
  {\bibfnamefont {R.}~\bibnamefont {Seshadri}}, \ and\ \bibinfo {author}
  {\bibfnamefont {S.~D.}\ \bibnamefont {Wilson}},\ }\bibfield  {title}
  {\enquote {\bibinfo {title} {Superconductivity in the {${\mathbb{Z}}_{2}$}
  kagome metal {KV$_3$Sb$_5$}},}\ }\href {\doibase
  10.1103/PhysRevMaterials.5.034801} {\bibfield  {journal} {\bibinfo  {journal}
  {Phys. Rev. Mater.}\ }\textbf {\bibinfo {volume} {5}},\ \bibinfo {pages}
  {034801} (\bibinfo {year} {2021}{\natexlab{a}})}\BibitemShut {NoStop}%
\bibitem [{\citenamefont {Yin}\ \emph {et~al.}(2021)\citenamefont {Yin},
  \citenamefont {Tu}, \citenamefont {Gong}, \citenamefont {Fu}, \citenamefont
  {Yan},\ and\ \citenamefont {Lei}}]{Yin2021}%
  \BibitemOpen
  \bibfield  {author} {\bibinfo {author} {\bibfnamefont {Q.}~\bibnamefont
  {Yin}}, \bibinfo {author} {\bibfnamefont {Z.}~\bibnamefont {Tu}}, \bibinfo
  {author} {\bibfnamefont {C.}~\bibnamefont {Gong}}, \bibinfo {author}
  {\bibfnamefont {Y.}~\bibnamefont {Fu}}, \bibinfo {author} {\bibfnamefont
  {S.}~\bibnamefont {Yan}}, \ and\ \bibinfo {author} {\bibfnamefont
  {H.}~\bibnamefont {Lei}},\ }\bibfield  {title} {\enquote {\bibinfo {title}
  {Superconductivity and normal-state properties of kagome metal
  {RbV$_3$Sb$_5$} single crystals},}\ }\href@noop {} {\bibfield  {journal}
  {\bibinfo  {journal} {Chin. Phys. Lett.}\ }\textbf {\bibinfo {volume} {38}},\
  \bibinfo {pages} {037403} (\bibinfo {year} {2021})}\BibitemShut {NoStop}%
\bibitem [{\citenamefont {Yu}\ \emph {et~al.}(2021{\natexlab{c}})\citenamefont
  {Yu}, \citenamefont {Ma}, \citenamefont {Zhuo}, \citenamefont {Liu},
  \citenamefont {Wen}, \citenamefont {Lei}, \citenamefont {Ying},\ and\
  \citenamefont {Chen}}]{Yu2021}%
  \BibitemOpen
  \bibfield  {author} {\bibinfo {author} {\bibfnamefont {F.}~\bibnamefont
  {Yu}}, \bibinfo {author} {\bibfnamefont {D.}~\bibnamefont {Ma}}, \bibinfo
  {author} {\bibfnamefont {W.}~\bibnamefont {Zhuo}}, \bibinfo {author}
  {\bibfnamefont {S.}~\bibnamefont {Liu}}, \bibinfo {author} {\bibfnamefont
  {X.}~\bibnamefont {Wen}}, \bibinfo {author} {\bibfnamefont {B.}~\bibnamefont
  {Lei}}, \bibinfo {author} {\bibfnamefont {J.}~\bibnamefont {Ying}}, \ and\
  \bibinfo {author} {\bibfnamefont {X.}~\bibnamefont {Chen}},\ }\bibfield
  {title} {\enquote {\bibinfo {title} {Unusual competition of superconductivity
  and charge-density-wave state in a compressed topological kagome metal},}\
  }\href@noop {} {\bibfield  {journal} {\bibinfo  {journal} {Nat. Commun.}\
  }\textbf {\bibinfo {volume} {12}},\ \bibinfo {pages} {3645} (\bibinfo {year}
  {2021}{\natexlab{c}})}\BibitemShut {NoStop}%
\bibitem [{\citenamefont {Wang}\ \emph
  {et~al.}(2021{\natexlab{b}})\citenamefont {Wang}, \citenamefont {Kong},
  \citenamefont {Shi}, \citenamefont {Pei}, \citenamefont {Wen}, \citenamefont
  {Gao}, \citenamefont {Zhao}, \citenamefont {Yin}, \citenamefont {Wu},
  \citenamefont {Li}, \citenamefont {Lei}, \citenamefont {Li}, \citenamefont
  {Chen}, \citenamefont {Yan},\ and\ \citenamefont {Qi}}]{Wang2021}%
  \BibitemOpen
  \bibfield  {author} {\bibinfo {author} {\bibfnamefont {Q.}~\bibnamefont
  {Wang}}, \bibinfo {author} {\bibfnamefont {P.}~\bibnamefont {Kong}}, \bibinfo
  {author} {\bibfnamefont {W.}~\bibnamefont {Shi}}, \bibinfo {author}
  {\bibfnamefont {C.}~\bibnamefont {Pei}}, \bibinfo {author} {\bibfnamefont
  {C.}~\bibnamefont {Wen}}, \bibinfo {author} {\bibfnamefont {L.}~\bibnamefont
  {Gao}}, \bibinfo {author} {\bibfnamefont {Y.}~\bibnamefont {Zhao}}, \bibinfo
  {author} {\bibfnamefont {Q.}~\bibnamefont {Yin}}, \bibinfo {author}
  {\bibfnamefont {Y.}~\bibnamefont {Wu}}, \bibinfo {author} {\bibfnamefont
  {G.}~\bibnamefont {Li}}, \bibinfo {author} {\bibfnamefont {H.}~\bibnamefont
  {Lei}}, \bibinfo {author} {\bibfnamefont {J.}~\bibnamefont {Li}}, \bibinfo
  {author} {\bibfnamefont {Y.}~\bibnamefont {Chen}}, \bibinfo {author}
  {\bibfnamefont {S.}~\bibnamefont {Yan}}, \ and\ \bibinfo {author}
  {\bibfnamefont {Y.}~\bibnamefont {Qi}},\ }\bibfield  {title} {\enquote
  {\bibinfo {title} {Charge density wave orders and enhanced superconductivity
  under pressure in the kagome metal {CsV$_3$Sb$_5$}},}\ }\href {\doibase
  https://doi.org/10.1002/adma.202102813} {\bibfield  {journal} {\bibinfo
  {journal} {Adv. Mater.}\ }\textbf {\bibinfo {volume} {33}},\ \bibinfo {pages}
  {2102813} (\bibinfo {year} {2021}{\natexlab{b}})}\BibitemShut {NoStop}%
\bibitem [{\citenamefont {Zhang}\ \emph {et~al.}(2021)\citenamefont {Zhang},
  \citenamefont {Chen}, \citenamefont {Zhou}, \citenamefont {Yuan},
  \citenamefont {Wang}, \citenamefont {Wang}, \citenamefont {Yang},
  \citenamefont {An}, \citenamefont {Zhang}, \citenamefont {Zhu}, \citenamefont
  {Zhou}, \citenamefont {Chen}, \citenamefont {Zhou},\ and\ \citenamefont
  {Yang}}]{Zhang2021}%
  \BibitemOpen
  \bibfield  {author} {\bibinfo {author} {\bibfnamefont {Z.}~\bibnamefont
  {Zhang}}, \bibinfo {author} {\bibfnamefont {Z.}~\bibnamefont {Chen}},
  \bibinfo {author} {\bibfnamefont {Y.}~\bibnamefont {Zhou}}, \bibinfo {author}
  {\bibfnamefont {Y.}~\bibnamefont {Yuan}}, \bibinfo {author} {\bibfnamefont
  {S.}~\bibnamefont {Wang}}, \bibinfo {author} {\bibfnamefont {J.}~\bibnamefont
  {Wang}}, \bibinfo {author} {\bibfnamefont {H.}~\bibnamefont {Yang}}, \bibinfo
  {author} {\bibfnamefont {C.}~\bibnamefont {An}}, \bibinfo {author}
  {\bibfnamefont {L.}~\bibnamefont {Zhang}}, \bibinfo {author} {\bibfnamefont
  {X.}~\bibnamefont {Zhu}}, \bibinfo {author} {\bibfnamefont {Y.}~\bibnamefont
  {Zhou}}, \bibinfo {author} {\bibfnamefont {X.}~\bibnamefont {Chen}}, \bibinfo
  {author} {\bibfnamefont {J.}~\bibnamefont {Zhou}}, \ and\ \bibinfo {author}
  {\bibfnamefont {Z.}~\bibnamefont {Yang}},\ }\bibfield  {title} {\enquote
  {\bibinfo {title} {Pressure-induced reemergence of superconductivity in the
  topological kagome metal {CsV$_3$Sb$_5$}},}\ }\href {\doibase
  10.1103/PhysRevB.103.224513} {\bibfield  {journal} {\bibinfo  {journal}
  {Phys. Rev. B}\ }\textbf {\bibinfo {volume} {103}},\ \bibinfo {pages}
  {224513} (\bibinfo {year} {2021})}\BibitemShut {NoStop}%
\bibitem [{\citenamefont {Wang}\ \emph
  {et~al.}(2021{\natexlab{c}})\citenamefont {Wang}, \citenamefont {Chen},
  \citenamefont {Yin}, \citenamefont {Ma}, \citenamefont {Pan}, \citenamefont
  {Yang}, \citenamefont {Ji}, \citenamefont {Wu}, \citenamefont {Shan},
  \citenamefont {Xu}, \citenamefont {Tu}, \citenamefont {Gong}, \citenamefont
  {Liu}, \citenamefont {Li}, \citenamefont {Uwatoko}, \citenamefont {Dong},
  \citenamefont {Lei}, \citenamefont {Sun},\ and\ \citenamefont
  {Cheng}}]{Wang2021a}%
  \BibitemOpen
  \bibfield  {author} {\bibinfo {author} {\bibfnamefont {N.~N.}\ \bibnamefont
  {Wang}}, \bibinfo {author} {\bibfnamefont {K.~Y.}\ \bibnamefont {Chen}},
  \bibinfo {author} {\bibfnamefont {Q.~W.}\ \bibnamefont {Yin}}, \bibinfo
  {author} {\bibfnamefont {Y.~N.~N.}\ \bibnamefont {Ma}}, \bibinfo {author}
  {\bibfnamefont {B.~Y.}\ \bibnamefont {Pan}}, \bibinfo {author} {\bibfnamefont
  {X.}~\bibnamefont {Yang}}, \bibinfo {author} {\bibfnamefont {X.~Y.}\
  \bibnamefont {Ji}}, \bibinfo {author} {\bibfnamefont {S.~L.}\ \bibnamefont
  {Wu}}, \bibinfo {author} {\bibfnamefont {P.~F.}\ \bibnamefont {Shan}},
  \bibinfo {author} {\bibfnamefont {S.~X.}\ \bibnamefont {Xu}}, \bibinfo
  {author} {\bibfnamefont {Z.~J.}\ \bibnamefont {Tu}}, \bibinfo {author}
  {\bibfnamefont {C.~S.}\ \bibnamefont {Gong}}, \bibinfo {author}
  {\bibfnamefont {G.~T.}\ \bibnamefont {Liu}}, \bibinfo {author} {\bibfnamefont
  {G.}~\bibnamefont {Li}}, \bibinfo {author} {\bibfnamefont {Y.}~\bibnamefont
  {Uwatoko}}, \bibinfo {author} {\bibfnamefont {X.~L.}\ \bibnamefont {Dong}},
  \bibinfo {author} {\bibfnamefont {H.~C.}\ \bibnamefont {Lei}}, \bibinfo
  {author} {\bibfnamefont {J.~P.}\ \bibnamefont {Sun}}, \ and\ \bibinfo
  {author} {\bibfnamefont {J.-G.}\ \bibnamefont {Cheng}},\ }\bibfield  {title}
  {\enquote {\bibinfo {title} {Competition between charge-density-wave and
  superconductivity in the kagome metal {RbV$_3$Sb$_5$}},}\ }\href {\doibase
  10.1103/PhysRevResearch.3.043018} {\bibfield  {journal} {\bibinfo  {journal}
  {Phys. Rev. Res.}\ }\textbf {\bibinfo {volume} {3}},\ \bibinfo {pages}
  {043018} (\bibinfo {year} {2021}{\natexlab{c}})}\BibitemShut {NoStop}%
\bibitem [{\citenamefont {Han}\ \emph {et~al.}(2017)\citenamefont {Han},
  \citenamefont {Xu}, \citenamefont {Botana}, \citenamefont {Xiao},
  \citenamefont {Wang}, \citenamefont {Yang}, \citenamefont {Chung},
  \citenamefont {Kanatzidis}, \citenamefont {Norman}, \citenamefont
  {Crabtree},\ and\ \citenamefont {Kwok}}]{Han2017}%
  \BibitemOpen
  \bibfield  {author} {\bibinfo {author} {\bibfnamefont {F.}~\bibnamefont
  {Han}}, \bibinfo {author} {\bibfnamefont {J.}~\bibnamefont {Xu}}, \bibinfo
  {author} {\bibfnamefont {A.~S.}\ \bibnamefont {Botana}}, \bibinfo {author}
  {\bibfnamefont {Z.~L.}\ \bibnamefont {Xiao}}, \bibinfo {author}
  {\bibfnamefont {Y.~L.}\ \bibnamefont {Wang}}, \bibinfo {author}
  {\bibfnamefont {W.~G.}\ \bibnamefont {Yang}}, \bibinfo {author}
  {\bibfnamefont {D.~Y.}\ \bibnamefont {Chung}}, \bibinfo {author}
  {\bibfnamefont {M.~G.}\ \bibnamefont {Kanatzidis}}, \bibinfo {author}
  {\bibfnamefont {M.~R.}\ \bibnamefont {Norman}}, \bibinfo {author}
  {\bibfnamefont {G.~W.}\ \bibnamefont {Crabtree}}, \ and\ \bibinfo {author}
  {\bibfnamefont {W.~K.}\ \bibnamefont {Kwok}},\ }\bibfield  {title} {\enquote
  {\bibinfo {title} {Separation of electron and hole dynamics in the semimetal
  {LaSb}},}\ }\href {\doibase 10.1103/PhysRevB.96.125112} {\bibfield  {journal}
  {\bibinfo  {journal} {Phys. Rev. B}\ }\textbf {\bibinfo {volume} {96}},\
  \bibinfo {pages} {125112} (\bibinfo {year} {2017})}\BibitemShut {NoStop}%
\bibitem [{\citenamefont {Hu}\ \emph {et~al.}(2018)\citenamefont {Hu},
  \citenamefont {Aulestia}, \citenamefont {Tse}, \citenamefont {Kuo},
  \citenamefont {Zhu}, \citenamefont {Lue}, \citenamefont {Lai},\ and\
  \citenamefont {Goh}}]{Hu2018}%
  \BibitemOpen
  \bibfield  {author} {\bibinfo {author} {\bibfnamefont {Y.~J.}\ \bibnamefont
  {Hu}}, \bibinfo {author} {\bibfnamefont {E.~I.~P.}\ \bibnamefont {Aulestia}},
  \bibinfo {author} {\bibfnamefont {K.~F.}\ \bibnamefont {Tse}}, \bibinfo
  {author} {\bibfnamefont {C.~N.}\ \bibnamefont {Kuo}}, \bibinfo {author}
  {\bibfnamefont {J.~Y.}\ \bibnamefont {Zhu}}, \bibinfo {author} {\bibfnamefont
  {C.~S.}\ \bibnamefont {Lue}}, \bibinfo {author} {\bibfnamefont {K.~T.}\
  \bibnamefont {Lai}}, \ and\ \bibinfo {author} {\bibfnamefont {S.~K.}\
  \bibnamefont {Goh}},\ }\bibfield  {title} {\enquote {\bibinfo {title}
  {Extremely large magnetoresistance and the complete determination of the
  {Fermi} surface topology in the semimetal {ScSb}},}\ }\href {\doibase
  10.1103/PhysRevB.98.035133} {\bibfield  {journal} {\bibinfo  {journal} {Phys.
  Rev. B}\ }\textbf {\bibinfo {volume} {98}},\ \bibinfo {pages} {035133}
  (\bibinfo {year} {2018})}\BibitemShut {NoStop}%
\bibitem [{\citenamefont {Rhodes}\ \emph {et~al.}(2017)\citenamefont {Rhodes},
  \citenamefont {Sch\"onemann}, \citenamefont {Aryal}, \citenamefont {Zhou},
  \citenamefont {Zhang}, \citenamefont {Kampert}, \citenamefont {Chiu},
  \citenamefont {Lai}, \citenamefont {Shimura}, \citenamefont {McCandless},
  \citenamefont {Chan}, \citenamefont {Paley}, \citenamefont {Lee},
  \citenamefont {Finke}, \citenamefont {Ruff}, \citenamefont {Das},
  \citenamefont {Manousakis},\ and\ \citenamefont {Balicas}}]{Rhodes2017}%
  \BibitemOpen
  \bibfield  {author} {\bibinfo {author} {\bibfnamefont {D.}~\bibnamefont
  {Rhodes}}, \bibinfo {author} {\bibfnamefont {R.}~\bibnamefont
  {Sch\"onemann}}, \bibinfo {author} {\bibfnamefont {N.}~\bibnamefont {Aryal}},
  \bibinfo {author} {\bibfnamefont {Q.}~\bibnamefont {Zhou}}, \bibinfo {author}
  {\bibfnamefont {Q.~R.}\ \bibnamefont {Zhang}}, \bibinfo {author}
  {\bibfnamefont {E.}~\bibnamefont {Kampert}}, \bibinfo {author} {\bibfnamefont
  {Y.-C.}\ \bibnamefont {Chiu}}, \bibinfo {author} {\bibfnamefont
  {Y.}~\bibnamefont {Lai}}, \bibinfo {author} {\bibfnamefont {Y.}~\bibnamefont
  {Shimura}}, \bibinfo {author} {\bibfnamefont {G.~T.}\ \bibnamefont
  {McCandless}}, \bibinfo {author} {\bibfnamefont {J.~Y.}\ \bibnamefont
  {Chan}}, \bibinfo {author} {\bibfnamefont {D.~W.}\ \bibnamefont {Paley}},
  \bibinfo {author} {\bibfnamefont {J.}~\bibnamefont {Lee}}, \bibinfo {author}
  {\bibfnamefont {A.~D.}\ \bibnamefont {Finke}}, \bibinfo {author}
  {\bibfnamefont {J.~P.~C.}\ \bibnamefont {Ruff}}, \bibinfo {author}
  {\bibfnamefont {S.}~\bibnamefont {Das}}, \bibinfo {author} {\bibfnamefont
  {E.}~\bibnamefont {Manousakis}}, \ and\ \bibinfo {author} {\bibfnamefont
  {L.}~\bibnamefont {Balicas}},\ }\bibfield  {title} {\enquote {\bibinfo
  {title} {Bulk {Fermi} surface of the {Weyl type-II} semimetallic candidate
  {$\ensuremath{\gamma}\ensuremath{-}{\mathrm{MoTe}}_{2}$}},}\ }\href {\doibase
  10.1103/PhysRevB.96.165134} {\bibfield  {journal} {\bibinfo  {journal} {Phys.
  Rev. B}\ }\textbf {\bibinfo {volume} {96}},\ \bibinfo {pages} {165134}
  (\bibinfo {year} {2017})}\BibitemShut {NoStop}%
\bibitem [{\citenamefont {Hu}\ \emph {et~al.}(2020)\citenamefont {Hu},
  \citenamefont {Yu}, \citenamefont {Lai}, \citenamefont {Sun}, \citenamefont
  {Balakirev}, \citenamefont {Zhang}, \citenamefont {Xie}, \citenamefont {Yip},
  \citenamefont {Aulestia}, \citenamefont {Jha}, \citenamefont {Higashinaka},
  \citenamefont {Matsuda}, \citenamefont {Yanase}, \citenamefont {Aoki},\ and\
  \citenamefont {Goh}}]{Hu2020}%
  \BibitemOpen
  \bibfield  {author} {\bibinfo {author} {\bibfnamefont {Y.~J.}\ \bibnamefont
  {Hu}}, \bibinfo {author} {\bibfnamefont {W.~C.}\ \bibnamefont {Yu}}, \bibinfo
  {author} {\bibfnamefont {K.~T.}\ \bibnamefont {Lai}}, \bibinfo {author}
  {\bibfnamefont {D.}~\bibnamefont {Sun}}, \bibinfo {author} {\bibfnamefont
  {F.~F.}\ \bibnamefont {Balakirev}}, \bibinfo {author} {\bibfnamefont
  {W.}~\bibnamefont {Zhang}}, \bibinfo {author} {\bibfnamefont {J.~Y.}\
  \bibnamefont {Xie}}, \bibinfo {author} {\bibfnamefont {K.~Y.}\ \bibnamefont
  {Yip}}, \bibinfo {author} {\bibfnamefont {E.~I.~P.}\ \bibnamefont
  {Aulestia}}, \bibinfo {author} {\bibfnamefont {R.}~\bibnamefont {Jha}},
  \bibinfo {author} {\bibfnamefont {R.}~\bibnamefont {Higashinaka}}, \bibinfo
  {author} {\bibfnamefont {T.~D.}\ \bibnamefont {Matsuda}}, \bibinfo {author}
  {\bibfnamefont {Y.}~\bibnamefont {Yanase}}, \bibinfo {author} {\bibfnamefont
  {Y.}~\bibnamefont {Aoki}}, \ and\ \bibinfo {author} {\bibfnamefont {S.~K.}\
  \bibnamefont {Goh}},\ }\bibfield  {title} {\enquote {\bibinfo {title}
  {Detection of hole pockets in the candidate type-{II Weyl} semimetal
  {${\mathrm{MoTe}}_{2}$} from {Shubnikov--de Haas} quantum oscillations},}\
  }\href {\doibase 10.1103/PhysRevLett.124.076402} {\bibfield  {journal}
  {\bibinfo  {journal} {Phys. Rev. Lett.}\ }\textbf {\bibinfo {volume} {124}},\
  \bibinfo {pages} {076402} (\bibinfo {year} {2020})}\BibitemShut {NoStop}%
\bibitem [{\citenamefont {Fu}\ \emph {et~al.}(2021)\citenamefont {Fu},
  \citenamefont {Zhao}, \citenamefont {Chen}, \citenamefont {Yin},
  \citenamefont {Tu}, \citenamefont {Gong}, \citenamefont {Xi}, \citenamefont
  {Zhu}, \citenamefont {Sun}, \citenamefont {Liu},\ and\ \citenamefont
  {Lei}}]{Fu2021}%
  \BibitemOpen
  \bibfield  {author} {\bibinfo {author} {\bibfnamefont {Y.}~\bibnamefont
  {Fu}}, \bibinfo {author} {\bibfnamefont {N.}~\bibnamefont {Zhao}}, \bibinfo
  {author} {\bibfnamefont {Z.}~\bibnamefont {Chen}}, \bibinfo {author}
  {\bibfnamefont {Q.}~\bibnamefont {Yin}}, \bibinfo {author} {\bibfnamefont
  {Z.}~\bibnamefont {Tu}}, \bibinfo {author} {\bibfnamefont {C.}~\bibnamefont
  {Gong}}, \bibinfo {author} {\bibfnamefont {C.}~\bibnamefont {Xi}}, \bibinfo
  {author} {\bibfnamefont {X.}~\bibnamefont {Zhu}}, \bibinfo {author}
  {\bibfnamefont {Y.}~\bibnamefont {Sun}}, \bibinfo {author} {\bibfnamefont
  {K.}~\bibnamefont {Liu}}, \ and\ \bibinfo {author} {\bibfnamefont
  {H.}~\bibnamefont {Lei}},\ }\bibfield  {title} {\enquote {\bibinfo {title}
  {Quantum transport evidence of topological band structures of kagome
  superconductor {CsV$_3$Sb$_5$}},}\ }\href {\doibase
  10.1103/PhysRevLett.127.207002} {\bibfield  {journal} {\bibinfo  {journal}
  {Phys. Rev. Lett.}\ }\textbf {\bibinfo {volume} {127}},\ \bibinfo {pages}
  {207002} (\bibinfo {year} {2021})}\BibitemShut {NoStop}%
\bibitem [{\citenamefont {Ortiz}\ \emph
  {et~al.}(2021{\natexlab{b}})\citenamefont {Ortiz}, \citenamefont {Teicher},
  \citenamefont {Kautzsch}, \citenamefont {Sarte}, \citenamefont {Ratcliff},
  \citenamefont {Harter}, \citenamefont {Ruff}, \citenamefont {Seshadri},\ and\
  \citenamefont {Wilson}}]{Ortiz2021}%
  \BibitemOpen
  \bibfield  {author} {\bibinfo {author} {\bibfnamefont {B.~R.}\ \bibnamefont
  {Ortiz}}, \bibinfo {author} {\bibfnamefont {S.~M.}\ \bibnamefont {Teicher}},
  \bibinfo {author} {\bibfnamefont {L.}~\bibnamefont {Kautzsch}}, \bibinfo
  {author} {\bibfnamefont {P.~M.}\ \bibnamefont {Sarte}}, \bibinfo {author}
  {\bibfnamefont {N.}~\bibnamefont {Ratcliff}}, \bibinfo {author}
  {\bibfnamefont {J.}~\bibnamefont {Harter}}, \bibinfo {author} {\bibfnamefont
  {J.~P.}\ \bibnamefont {Ruff}}, \bibinfo {author} {\bibfnamefont
  {R.}~\bibnamefont {Seshadri}}, \ and\ \bibinfo {author} {\bibfnamefont
  {S.~D.}\ \bibnamefont {Wilson}},\ }\bibfield  {title} {\enquote {\bibinfo
  {title} {Fermi surface mapping and the nature of charge-density-wave order in
  the kagome superconductor {CsV$_3$Sb$_5$}},}\ }\href@noop {} {\bibfield
  {journal} {\bibinfo  {journal} {Phys. Rev. X}\ }\textbf {\bibinfo {volume}
  {11}},\ \bibinfo {pages} {041030} (\bibinfo {year}
  {2021}{\natexlab{b}})}\BibitemShut {NoStop}%
\bibitem [{\citenamefont {Gan}\ \emph {et~al.}(2021)\citenamefont {Gan},
  \citenamefont {Xia}, \citenamefont {Zhang}, \citenamefont {Yang},
  \citenamefont {Mi}, \citenamefont {Wang}, \citenamefont {Chai}, \citenamefont
  {Guo}, \citenamefont {Zhou},\ and\ \citenamefont {He}}]{Gan2021}%
  \BibitemOpen
  \bibfield  {author} {\bibinfo {author} {\bibfnamefont {Y.}~\bibnamefont
  {Gan}}, \bibinfo {author} {\bibfnamefont {W.}~\bibnamefont {Xia}}, \bibinfo
  {author} {\bibfnamefont {L.}~\bibnamefont {Zhang}}, \bibinfo {author}
  {\bibfnamefont {K.}~\bibnamefont {Yang}}, \bibinfo {author} {\bibfnamefont
  {X.}~\bibnamefont {Mi}}, \bibinfo {author} {\bibfnamefont {A.}~\bibnamefont
  {Wang}}, \bibinfo {author} {\bibfnamefont {Y.}~\bibnamefont {Chai}}, \bibinfo
  {author} {\bibfnamefont {Y.}~\bibnamefont {Guo}}, \bibinfo {author}
  {\bibfnamefont {X.}~\bibnamefont {Zhou}}, \ and\ \bibinfo {author}
  {\bibfnamefont {M.}~\bibnamefont {He}},\ }\bibfield  {title} {\enquote
  {\bibinfo {title} {Magneto-{Seebeck} effect and ambipolar {Nernst} effect in
  the {CsV$_3$Sb$_5$} superconductor},}\ }\href@noop {} {\bibfield  {journal}
  {\bibinfo  {journal} {Phys. Rev. B}\ }\textbf {\bibinfo {volume} {104}},\
  \bibinfo {pages} {L180508} (\bibinfo {year} {2021})}\BibitemShut {NoStop}%
\bibitem [{\citenamefont {Chen}\ \emph {et~al.}(2022)\citenamefont {Chen},
  \citenamefont {He}, \citenamefont {Yao}, \citenamefont {Pan}, \citenamefont
  {Lin}, \citenamefont {Schnelle}, \citenamefont {Sun}, \citenamefont {Gooth},
  \citenamefont {Taillefer},\ and\ \citenamefont {Felser}}]{Chen2022}%
  \BibitemOpen
  \bibfield  {author} {\bibinfo {author} {\bibfnamefont {D.}~\bibnamefont
  {Chen}}, \bibinfo {author} {\bibfnamefont {B.}~\bibnamefont {He}}, \bibinfo
  {author} {\bibfnamefont {M.}~\bibnamefont {Yao}}, \bibinfo {author}
  {\bibfnamefont {Y.}~\bibnamefont {Pan}}, \bibinfo {author} {\bibfnamefont
  {H.}~\bibnamefont {Lin}}, \bibinfo {author} {\bibfnamefont {W.}~\bibnamefont
  {Schnelle}}, \bibinfo {author} {\bibfnamefont {Y.}~\bibnamefont {Sun}},
  \bibinfo {author} {\bibfnamefont {J.}~\bibnamefont {Gooth}}, \bibinfo
  {author} {\bibfnamefont {L.}~\bibnamefont {Taillefer}}, \ and\ \bibinfo
  {author} {\bibfnamefont {C.}~\bibnamefont {Felser}},\ }\bibfield  {title}
  {\enquote {\bibinfo {title} {Anomalous thermoelectric effects and quantum
  oscillations in the kagome metal {CsV$_3$Sb$_5$}},}\ }\href@noop {}
  {\bibfield  {journal} {\bibinfo  {journal} {Phys. Rev. B}\ }\textbf {\bibinfo
  {volume} {105}},\ \bibinfo {pages} {L201109} (\bibinfo {year}
  {2022})}\BibitemShut {NoStop}%
\bibitem [{\citenamefont {Huang}\ \emph {et~al.}(2022)\citenamefont {Huang},
  \citenamefont {Guo}, \citenamefont {Putzke}, \citenamefont {Gutierrez-Amigo},
  \citenamefont {Sun}, \citenamefont {Vergniory}, \citenamefont {Errea},
  \citenamefont {Chen}, \citenamefont {Felser},\ and\ \citenamefont
  {Moll}}]{Huang2022}%
  \BibitemOpen
  \bibfield  {author} {\bibinfo {author} {\bibfnamefont {X.}~\bibnamefont
  {Huang}}, \bibinfo {author} {\bibfnamefont {C.}~\bibnamefont {Guo}}, \bibinfo
  {author} {\bibfnamefont {C.}~\bibnamefont {Putzke}}, \bibinfo {author}
  {\bibfnamefont {M.}~\bibnamefont {Gutierrez-Amigo}}, \bibinfo {author}
  {\bibfnamefont {Y.}~\bibnamefont {Sun}}, \bibinfo {author} {\bibfnamefont
  {M.~G.}\ \bibnamefont {Vergniory}}, \bibinfo {author} {\bibfnamefont
  {I.}~\bibnamefont {Errea}}, \bibinfo {author} {\bibfnamefont
  {D.}~\bibnamefont {Chen}}, \bibinfo {author} {\bibfnamefont {C.}~\bibnamefont
  {Felser}}, \ and\ \bibinfo {author} {\bibfnamefont {P.~J.~W.}\ \bibnamefont
  {Moll}},\ }\bibfield  {title} {\enquote {\bibinfo {title} {Three-dimensional
  {Fermi} surfaces from charge order in layered {CsV$_3$Sb$_5$}},}\ }\href
  {\doibase 10.1103/PhysRevB.106.064510} {\bibfield  {journal} {\bibinfo
  {journal} {Phys. Rev. B}\ }\textbf {\bibinfo {volume} {106}},\ \bibinfo
  {pages} {064510} (\bibinfo {year} {2022})}\BibitemShut {NoStop}%
\bibitem [{\citenamefont {Shrestha}\ \emph {et~al.}(2022)\citenamefont
  {Shrestha}, \citenamefont {Chapai}, \citenamefont {Pokharel}, \citenamefont
  {Miertschin}, \citenamefont {Nguyen}, \citenamefont {Zhou}, \citenamefont
  {Chung}, \citenamefont {Kanatzidis}, \citenamefont {Mitchell}, \citenamefont
  {Welp}, \citenamefont {Popovi\ifmmode~\acute{c}\else \'{c}\fi{}},
  \citenamefont {Graf}, \citenamefont {Lorenz},\ and\ \citenamefont
  {Kwok}}]{Shrestha2022}%
  \BibitemOpen
  \bibfield  {author} {\bibinfo {author} {\bibfnamefont {K.}~\bibnamefont
  {Shrestha}}, \bibinfo {author} {\bibfnamefont {R.}~\bibnamefont {Chapai}},
  \bibinfo {author} {\bibfnamefont {B.~K.}\ \bibnamefont {Pokharel}}, \bibinfo
  {author} {\bibfnamefont {D.}~\bibnamefont {Miertschin}}, \bibinfo {author}
  {\bibfnamefont {T.}~\bibnamefont {Nguyen}}, \bibinfo {author} {\bibfnamefont
  {X.}~\bibnamefont {Zhou}}, \bibinfo {author} {\bibfnamefont {D.~Y.}\
  \bibnamefont {Chung}}, \bibinfo {author} {\bibfnamefont {M.~G.}\ \bibnamefont
  {Kanatzidis}}, \bibinfo {author} {\bibfnamefont {J.~F.}\ \bibnamefont
  {Mitchell}}, \bibinfo {author} {\bibfnamefont {U.}~\bibnamefont {Welp}},
  \bibinfo {author} {\bibfnamefont {D.}~\bibnamefont
  {Popovi\ifmmode~\acute{c}\else \'{c}\fi{}}}, \bibinfo {author} {\bibfnamefont
  {D.~E.}\ \bibnamefont {Graf}}, \bibinfo {author} {\bibfnamefont
  {B.}~\bibnamefont {Lorenz}}, \ and\ \bibinfo {author} {\bibfnamefont {W.~K.}\
  \bibnamefont {Kwok}},\ }\bibfield  {title} {\enquote {\bibinfo {title}
  {Nontrivial {Fermi} surface topology of the kagome superconductor
  {CsV$_3$Sb$_5$} probed by {de Haas--van Alphen} oscillations},}\ }\href
  {\doibase 10.1103/PhysRevB.105.024508} {\bibfield  {journal} {\bibinfo
  {journal} {Phys. Rev. B}\ }\textbf {\bibinfo {volume} {105}},\ \bibinfo
  {pages} {024508} (\bibinfo {year} {2022})}\BibitemShut {NoStop}%
\bibitem [{\citenamefont {Xie}\ \emph {et~al.}(2021)\citenamefont {Xie},
  \citenamefont {Liu}, \citenamefont {Zhang}, \citenamefont {Wong},
  \citenamefont {Zhou}, \citenamefont {Zhao}, \citenamefont {Wang},
  \citenamefont {Lai},\ and\ \citenamefont {Goh}}]{Xie2021}%
  \BibitemOpen
  \bibfield  {author} {\bibinfo {author} {\bibfnamefont {J.}~\bibnamefont
  {Xie}}, \bibinfo {author} {\bibfnamefont {X.}~\bibnamefont {Liu}}, \bibinfo
  {author} {\bibfnamefont {W.}~\bibnamefont {Zhang}}, \bibinfo {author}
  {\bibfnamefont {S.~M.}\ \bibnamefont {Wong}}, \bibinfo {author}
  {\bibfnamefont {X.}~\bibnamefont {Zhou}}, \bibinfo {author} {\bibfnamefont
  {Y.}~\bibnamefont {Zhao}}, \bibinfo {author} {\bibfnamefont {S.}~\bibnamefont
  {Wang}}, \bibinfo {author} {\bibfnamefont {K.~T.}\ \bibnamefont {Lai}}, \
  and\ \bibinfo {author} {\bibfnamefont {S.~K.}\ \bibnamefont {Goh}},\
  }\bibfield  {title} {\enquote {\bibinfo {title} {Fragile pressure-induced
  magnetism in {FeSe} superconductors with a thickness reduction},}\
  }\href@noop {} {\bibfield  {journal} {\bibinfo  {journal} {Nano Lett.}\
  }\textbf {\bibinfo {volume} {21}},\ \bibinfo {pages} {9310} (\bibinfo {year}
  {2021})}\BibitemShut {NoStop}%
\bibitem [{SUP()}]{SUPP}%
  \BibitemOpen
  \href@noop {} {}\bibinfo {note} {See Supplemental Material for (i) details of
  density functional theory calculations, (ii) temperature dependence of
  electrical resistivity for a thin flake of CsV$_3$Sb$_5$, (iii) calculated
  Fermi surfaces for the distorted CsV$_3$Sb$_5$~($2 \times 2 \times 4$
  superstructure). The Supplemental Material includes additional
  Refs.~[\onlinecite{schwarz2003solid,perdew1996generalized,julian2012numerical}]}\BibitemShut
  {NoStop}%
  \bibitem [{\citenamefont {Schwarz}\ and\ \citenamefont
  {Blaha}(2003)}]{schwarz2003solid}%
  \BibitemOpen
  \bibfield  {author} {\bibinfo {author} {\bibfnamefont {K.}~\bibnamefont
  {Schwarz}}\ and\ \bibinfo {author} {\bibfnamefont {P.}~\bibnamefont
  {Blaha}},\ }\bibfield  {title} {\enquote {\bibinfo {title} {Solid state
  calculations using {WIEN2k}},}\ }\href@noop {} {\bibfield  {journal}
  {\bibinfo  {journal} {Comput. Mater. Sci.}\ }\textbf {\bibinfo {volume}
  {28}},\ \bibinfo {pages} {259} (\bibinfo {year} {2003})}\BibitemShut
  {NoStop}%
\bibitem [{\citenamefont {Perdew}\ \emph {et~al.}(1996)\citenamefont {Perdew},
  \citenamefont {Burke},\ and\ \citenamefont
  {Ernzerhof}}]{perdew1996generalized}%
  \BibitemOpen
  \bibfield  {author} {\bibinfo {author} {\bibfnamefont {J.~P.}\ \bibnamefont
  {Perdew}}, \bibinfo {author} {\bibfnamefont {K.}~\bibnamefont {Burke}}, \
  and\ \bibinfo {author} {\bibfnamefont {M.}~\bibnamefont {Ernzerhof}},\
  }\bibfield  {title} {\enquote {\bibinfo {title} {Generalized gradient
  approximation made simple},}\ }\href@noop {} {\bibfield  {journal} {\bibinfo
  {journal} {Phys. Rev. Lett.}\ }\textbf {\bibinfo {volume} {77}},\ \bibinfo
  {pages} {3865} (\bibinfo {year} {1996})}\BibitemShut {NoStop}%
\bibitem [{\citenamefont {Rourke}\ and\ \citenamefont
  {Julian}(2012)}]{julian2012numerical}%
  \BibitemOpen
  \bibfield  {author} {\bibinfo {author} {\bibfnamefont {P.}~\bibnamefont
  {Rourke}}\ and\ \bibinfo {author} {\bibfnamefont {S.}~\bibnamefont
  {Julian}},\ }\bibfield  {title} {\enquote {\bibinfo {title} {Numerical
  extraction of de {Haas}–van {Alphen} frequencies from calculated band
  energies},}\ }\href {\doibase https://doi.org/10.1016/j.cpc.2011.10.015}
  {\bibfield  {journal} {\bibinfo  {journal} {Comput. Phys. Commun.}\ }\textbf
  {\bibinfo {volume} {183}},\ \bibinfo {pages} {324} (\bibinfo {year}
  {2012})}\BibitemShut {NoStop}%
\bibitem [{\citenamefont {Song}\ \emph {et~al.}(2021)\citenamefont {Song},
  \citenamefont {Ying}, \citenamefont {Chen}, \citenamefont {Han},
  \citenamefont {Wu}, \citenamefont {Schnyder}, \citenamefont {Huang},
  \citenamefont {Guo},\ and\ \citenamefont {Chen}}]{Song2021}%
  \BibitemOpen
  \bibfield  {author} {\bibinfo {author} {\bibfnamefont {Y.}~\bibnamefont
  {Song}}, \bibinfo {author} {\bibfnamefont {T.}~\bibnamefont {Ying}}, \bibinfo
  {author} {\bibfnamefont {X.}~\bibnamefont {Chen}}, \bibinfo {author}
  {\bibfnamefont {X.}~\bibnamefont {Han}}, \bibinfo {author} {\bibfnamefont
  {X.}~\bibnamefont {Wu}}, \bibinfo {author} {\bibfnamefont {A.~P.}\
  \bibnamefont {Schnyder}}, \bibinfo {author} {\bibfnamefont {Y.}~\bibnamefont
  {Huang}}, \bibinfo {author} {\bibfnamefont {J.-G.}\ \bibnamefont {Guo}}, \
  and\ \bibinfo {author} {\bibfnamefont {X.}~\bibnamefont {Chen}},\ }\bibfield
  {title} {\enquote {\bibinfo {title} {Competition of superconductivity and
  charge density wave in selective oxidized {CsV$_3$Sb$_5$} thin flakes},}\
  }\href {\doibase 10.1103/PhysRevLett.127.237001} {\bibfield  {journal}
  {\bibinfo  {journal} {Phys. Rev. Lett.}\ }\textbf {\bibinfo {volume} {127}},\
  \bibinfo {pages} {237001} (\bibinfo {year} {2021})}\BibitemShut {NoStop}%
\bibitem [{\citenamefont {Qu}\ \emph {et~al.}(2010)\citenamefont {Qu},
  \citenamefont {Hor}, \citenamefont {Xiong}, \citenamefont {Cava},\ and\
  \citenamefont {Ong}}]{Qu2010}%
  \BibitemOpen
  \bibfield  {author} {\bibinfo {author} {\bibfnamefont {D.-X.}\ \bibnamefont
  {Qu}}, \bibinfo {author} {\bibfnamefont {Y.~S.}\ \bibnamefont {Hor}},
  \bibinfo {author} {\bibfnamefont {J.}~\bibnamefont {Xiong}}, \bibinfo
  {author} {\bibfnamefont {R.~J.}\ \bibnamefont {Cava}}, \ and\ \bibinfo
  {author} {\bibfnamefont {N.~P.}\ \bibnamefont {Ong}},\ }\bibfield  {title}
  {\enquote {\bibinfo {title} {Quantum oscillations and {Hall} anomaly of
  surface states in the topological insulator {Bi$_2$Te$_3$}},}\ }\href@noop {}
  {\bibfield  {journal} {\bibinfo  {journal} {Science}\ }\textbf {\bibinfo
  {volume} {329}},\ \bibinfo {pages} {821} (\bibinfo {year}
  {2010})}\BibitemShut {NoStop}%
\bibitem [{\citenamefont {Taskin}\ \emph {et~al.}(2011)\citenamefont {Taskin},
  \citenamefont {Ren}, \citenamefont {Sasaki}, \citenamefont {Segawa},\ and\
  \citenamefont {Ando}}]{Taskin2011}%
  \BibitemOpen
  \bibfield  {author} {\bibinfo {author} {\bibfnamefont {A.}~\bibnamefont
  {Taskin}}, \bibinfo {author} {\bibfnamefont {Z.}~\bibnamefont {Ren}},
  \bibinfo {author} {\bibfnamefont {S.}~\bibnamefont {Sasaki}}, \bibinfo
  {author} {\bibfnamefont {K.}~\bibnamefont {Segawa}}, \ and\ \bibinfo {author}
  {\bibfnamefont {Y.}~\bibnamefont {Ando}},\ }\bibfield  {title} {\enquote
  {\bibinfo {title} {Observation of {Dirac} holes and electrons in a
  topological insulator},}\ }\href@noop {} {\bibfield  {journal} {\bibinfo
  {journal} {Phys. Rev. Lett.}\ }\textbf {\bibinfo {volume} {107}},\ \bibinfo
  {pages} {016801} (\bibinfo {year} {2011})}\BibitemShut {NoStop}%
\bibitem [{\citenamefont {Analytis}\ \emph {et~al.}(2010)\citenamefont
  {Analytis}, \citenamefont {McDonald}, \citenamefont {Riggs}, \citenamefont
  {Chu}, \citenamefont {Boebinger},\ and\ \citenamefont
  {Fisher}}]{Analytis2010}%
  \BibitemOpen
  \bibfield  {author} {\bibinfo {author} {\bibfnamefont {J.~G.}\ \bibnamefont
  {Analytis}}, \bibinfo {author} {\bibfnamefont {R.~D.}\ \bibnamefont
  {McDonald}}, \bibinfo {author} {\bibfnamefont {S.~C.}\ \bibnamefont {Riggs}},
  \bibinfo {author} {\bibfnamefont {J.-H.}\ \bibnamefont {Chu}}, \bibinfo
  {author} {\bibfnamefont {G.}~\bibnamefont {Boebinger}}, \ and\ \bibinfo
  {author} {\bibfnamefont {I.~R.}\ \bibnamefont {Fisher}},\ }\bibfield  {title}
  {\enquote {\bibinfo {title} {Two-dimensional surface state in the quantum
  limit of a topological insulator},}\ }\href@noop {} {\bibfield  {journal}
  {\bibinfo  {journal} {Nat. Phys.}\ }\textbf {\bibinfo {volume} {6}},\
  \bibinfo {pages} {960} (\bibinfo {year} {2010})}\BibitemShut {NoStop}%
\bibitem [{\citenamefont {Kim}\ \emph {et~al.}(2020)\citenamefont {Kim},
  \citenamefont {Hwang}, \citenamefont {Kim}, \citenamefont {Hou},
  \citenamefont {Yu}, \citenamefont {Sim},\ and\ \citenamefont
  {Doh}}]{Kim2020}%
  \BibitemOpen
  \bibfield  {author} {\bibinfo {author} {\bibfnamefont {H.-S.}\ \bibnamefont
  {Kim}}, \bibinfo {author} {\bibfnamefont {T.-H.}\ \bibnamefont {Hwang}},
  \bibinfo {author} {\bibfnamefont {N.-H.}\ \bibnamefont {Kim}}, \bibinfo
  {author} {\bibfnamefont {Y.}~\bibnamefont {Hou}}, \bibinfo {author}
  {\bibfnamefont {D.}~\bibnamefont {Yu}}, \bibinfo {author} {\bibfnamefont
  {H.-S.}\ \bibnamefont {Sim}}, \ and\ \bibinfo {author} {\bibfnamefont
  {Y.-J.}\ \bibnamefont {Doh}},\ }\bibfield  {title} {\enquote {\bibinfo
  {title} {Adjustable quantum interference oscillations in {Sb}-doped
  {Bi$_2$Se$_3$} topological insulator nanoribbons},}\ }\href@noop {}
  {\bibfield  {journal} {\bibinfo  {journal} {ACS nano}\ }\textbf {\bibinfo
  {volume} {14}},\ \bibinfo {pages} {14118} (\bibinfo {year}
  {2020})}\BibitemShut {NoStop}%
\bibitem [{foo()}]{footnote1}%
  \BibitemOpen
  \href@noop {} {}\bibinfo {note} {We take the 2167 T frequency as an example.
  For the ease of argument, we assume a circular orbit in the $k_x$-$k_y$
  plane. The 2167 T frequency follows $F(\theta)=F(0^\circ)/\cos\theta$ well up
  to 23.6$^\circ$ and at 23.6$^\circ$, the ‘$k_z$-height’ of the extremal
  orbit associated with the 2167 T frequency is around 0.224 $\rm \AA^{-1}$,
  which is much larger than the height of the first Brillouin zone (0.168 $\rm
  \AA^{-1}$ for the $2\times2\times4$ superstructure). At 23.6$^\circ$, the
  quantum oscillation orbit has already traveled beyond the first Brillouin
  zone and the Fermi surface is never closed. Hence, even with data up to
  23.6$^\circ$ only, we can safely conclude the quasi-2D nature of this Fermi
  surface sheet.}\BibitemShut {Stop}%
\bibitem [{\citenamefont {Tan}\ \emph {et~al.}(2021)\citenamefont {Tan},
  \citenamefont {Liu}, \citenamefont {Wang},\ and\ \citenamefont
  {Yan}}]{Tan2021}%
  \BibitemOpen
  \bibfield  {author} {\bibinfo {author} {\bibfnamefont {H.}~\bibnamefont
  {Tan}}, \bibinfo {author} {\bibfnamefont {Y.}~\bibnamefont {Liu}}, \bibinfo
  {author} {\bibfnamefont {Z.}~\bibnamefont {Wang}}, \ and\ \bibinfo {author}
  {\bibfnamefont {B.}~\bibnamefont {Yan}},\ }\bibfield  {title} {\enquote
  {\bibinfo {title} {Charge density waves and electronic properties of
  superconducting kagome metals},}\ }\href {\doibase
  10.1103/PhysRevLett.127.046401} {\bibfield  {journal} {\bibinfo  {journal}
  {Phys. Rev. Lett.}\ }\textbf {\bibinfo {volume} {127}},\ \bibinfo {pages}
  {046401} (\bibinfo {year} {2021})}\BibitemShut {NoStop}%
\bibitem [{\citenamefont {Shoenberg}(1984)}]{Shoenberg_book}%
  \BibitemOpen
  \bibfield  {author} {\bibinfo {author} {\bibfnamefont {D.}~\bibnamefont
  {Shoenberg}},\ }\href@noop {} {\emph {\bibinfo {title} {Magnetic Oscillations
  in Metals.}}}\ (\bibinfo  {publisher} {Cambridge Univ. Press},\ \bibinfo
  {year} {1984})\BibitemShut {NoStop}%
\bibitem [{\citenamefont {Friedemann}\ \emph {et~al.}(2013)\citenamefont
  {Friedemann}, \citenamefont {Goh}, \citenamefont {Rourke}, \citenamefont
  {Reiss}, \citenamefont {Sutherland}, \citenamefont {Grosche}, \citenamefont
  {Zwicknagl},\ and\ \citenamefont {Fisk}}]{Friedemann2013}%
  \BibitemOpen
  \bibfield  {author} {\bibinfo {author} {\bibfnamefont {S.}~\bibnamefont
  {Friedemann}}, \bibinfo {author} {\bibfnamefont {S.~K.}\ \bibnamefont {Goh}},
  \bibinfo {author} {\bibfnamefont {P.~M.~C.}\ \bibnamefont {Rourke}}, \bibinfo
  {author} {\bibfnamefont {P.}~\bibnamefont {Reiss}}, \bibinfo {author}
  {\bibfnamefont {M.~L.}\ \bibnamefont {Sutherland}}, \bibinfo {author}
  {\bibfnamefont {F.~M.}\ \bibnamefont {Grosche}}, \bibinfo {author}
  {\bibfnamefont {G.}~\bibnamefont {Zwicknagl}}, \ and\ \bibinfo {author}
  {\bibfnamefont {Z.}~\bibnamefont {Fisk}},\ }\bibfield  {title} {\enquote
  {\bibinfo {title} {Electronic structure of {LuRh$_2$Si$_2$}: `small' fermi
  surface reference to {YbRh$_2$Si$_2$}},}\ }\href {\doibase
  10.1088/1367-2630/15/9/093014} {\bibfield  {journal} {\bibinfo  {journal}
  {New J. Phys.}\ }\textbf {\bibinfo {volume} {15}},\ \bibinfo {pages} {093014}
  (\bibinfo {year} {2013})}\BibitemShut {NoStop}%
\bibitem [{\citenamefont {Luo}\ \emph {et~al.}(2021)\citenamefont {Luo},
  \citenamefont {Peng}, \citenamefont {Teicher}, \citenamefont {Huai},
  \citenamefont {Hu}, \citenamefont {Ortiz}, \citenamefont {Wei}, \citenamefont
  {Shen}, \citenamefont {Ou}, \citenamefont {Wang}, \citenamefont {Miao},
  \citenamefont {Guo}, \citenamefont {Shi}, \citenamefont {Wilson},\ and\
  \citenamefont {He}}]{Luo2021}%
  \BibitemOpen
  \bibfield  {author} {\bibinfo {author} {\bibfnamefont {Y.}~\bibnamefont
  {Luo}}, \bibinfo {author} {\bibfnamefont {S.}~\bibnamefont {Peng}}, \bibinfo
  {author} {\bibfnamefont {S.~M.~L.}\ \bibnamefont {Teicher}}, \bibinfo
  {author} {\bibfnamefont {L.}~\bibnamefont {Huai}}, \bibinfo {author}
  {\bibfnamefont {Y.}~\bibnamefont {Hu}}, \bibinfo {author} {\bibfnamefont
  {B.~R.}\ \bibnamefont {Ortiz}}, \bibinfo {author} {\bibfnamefont
  {Z.}~\bibnamefont {Wei}}, \bibinfo {author} {\bibfnamefont {J.}~\bibnamefont
  {Shen}}, \bibinfo {author} {\bibfnamefont {Z.}~\bibnamefont {Ou}}, \bibinfo
  {author} {\bibfnamefont {B.}~\bibnamefont {Wang}}, \bibinfo {author}
  {\bibfnamefont {Y.}~\bibnamefont {Miao}}, \bibinfo {author} {\bibfnamefont
  {M.}~\bibnamefont {Guo}}, \bibinfo {author} {\bibfnamefont {M.}~\bibnamefont
  {Shi}}, \bibinfo {author} {\bibfnamefont {S.~D.}\ \bibnamefont {Wilson}}, \
  and\ \bibinfo {author} {\bibfnamefont {J.~F.}\ \bibnamefont {He}},\
  }\bibfield  {title} {\enquote {\bibinfo {title} {Distinct band
  reconstructions in kagome superconductor {CsV$_3$Sb$_5$}},}\ }\href@noop {}
  {\bibfield  {journal} {\bibinfo  {journal} {arXiv:2106.01248}\ } (\bibinfo
  {year} {2021})}\BibitemShut {NoStop}%
\bibitem [{\citenamefont {Chapai}\ \emph {et~al.}(2022)\citenamefont {Chapai},
  \citenamefont {Leroux}, \citenamefont {Oliviero}, \citenamefont {Vignolles},
  \citenamefont {Smylie}, \citenamefont {Chung}, \citenamefont {Kanatzidis},
  \citenamefont {Kwok}, \citenamefont {Mitchell},\ and\ \citenamefont
  {Welp}}]{Chapai2022}%
  \BibitemOpen
  \bibfield  {author} {\bibinfo {author} {\bibfnamefont {R.}~\bibnamefont
  {Chapai}}, \bibinfo {author} {\bibfnamefont {M.}~\bibnamefont {Leroux}},
  \bibinfo {author} {\bibfnamefont {V.}~\bibnamefont {Oliviero}}, \bibinfo
  {author} {\bibfnamefont {D.}~\bibnamefont {Vignolles}}, \bibinfo {author}
  {\bibfnamefont {M.}~\bibnamefont {Smylie}}, \bibinfo {author} {\bibfnamefont
  {D.}~\bibnamefont {Chung}}, \bibinfo {author} {\bibfnamefont
  {M.}~\bibnamefont {Kanatzidis}}, \bibinfo {author} {\bibfnamefont {W.-K.}\
  \bibnamefont {Kwok}}, \bibinfo {author} {\bibfnamefont {J.}~\bibnamefont
  {Mitchell}}, \ and\ \bibinfo {author} {\bibfnamefont {U.}~\bibnamefont
  {Welp}},\ }\bibfield  {title} {\enquote {\bibinfo {title} {Magnetic breakdown
  and topology in the kagome superconductor {CsV$_3$Sb$_5$} under high magnetic
  field},}\ }\href@noop {} {\bibfield  {journal} {\bibinfo  {journal}
  {arXiv:2208.05523}\ } (\bibinfo {year} {2022})}\BibitemShut {NoStop}%
\end{thebibliography}


\providecommand{\noopsort}[1]{}\providecommand{\singleletter}[1]{#1}%

\end{document}